\documentclass[12pt]{article}

\usepackage{scicite}
\usepackage{times}
\usepackage[normalem]{ulem}

\usepackage{epsfig}
\usepackage{xspace}
\usepackage{amssymb}
\usepackage{amsmath}
\usepackage{lineno}
\usepackage{color}
\usepackage{url}
\usepackage{hyperref}
\usepackage[utf8]{inputenc}
\usepackage{multirow}

\def\aap{Astron. \& Astrophys.}                
\newcommand{\aj}{Astron. J.}
\newcommand{\aapr}{Astron. Astrophys. Rev.}

\newcommand{\apj}{Astrophys. J.}
\newcommand{\apjl}{Astrophys. J. Lett.}
\newcommand{\mnras}{Mon. Not. R. Astron. Soc.}
\newcommand{\pasj}{Publ. Astron. Soc. Japan}

\newcommand{\procspie}{Proc. SPIE}

\newcommand{\ssr}{Space Sci. Rev.}

\newcommand\xspec{{\sc xspec}\xspace}
\newcommand\ixpeobssim{{\sc ixpeobssim} \xspace}

\newcommand{\degr}{\hbox{$^{\circ}$}}

\newcommand\fdg{\mbox{$.\!\!^\circ$}}%

\newcommand{\info}[2]{{{#2}$^{#1}$,}}
\newcommand{\infoe}[2]{{and {#2}$^{#1}$}}
\newcommand{\inst}[2]{{$^{#1}${#2}\newline}}
\begin{document} 
\sloppy 
\begin{center}
{\LARGE Polarized x-rays constrain the disk-jet geometry in the black hole x-ray binary Cygnus X-1}\\[4ex]
{\large
\info{1,*}{Henric Krawczynski}
\info{2,*}{Fabio Muleri}
\info{3,*}{Michal Dovčiak}
\info{4,5,6,*}{Alexandra Veledina}
\info{1}{Nicole Rodriguez Cavero}
\info{3}{Jiri Svoboda}
\info{7}{Adam Ingram}
\info{8}{Giorgio Matt}
\info{9}{Javier A. Garcia}
\info{4}{Vladislav Loktev}
\info{10,11,12}{Michela Negro}
\info{4,6}{Juri Poutanen}
\info{13}{Takao Kitaguchi}
\info{14,3,15}{Jakub Podgorný}
\info{2}{John Rankin}
\info{16}{Wenda Z{h}ang}
\info{4}{Andrei Berdyugin}
\info{17,18,19}{Svetlana V. Berdyugina}
\info{8}{Stefano Bianchi}
\info{20,21}{Dmitry Blinov}
\info{2}{Fiamma Capitanio}
\info{22}{Niccolò Di Lalla}
\info{23}{Paul Draghis}
\info{2}{Sergio Fabiani}
\info{24}{Masato Kagitani}
\info{4}{Vadim Kravtsov}
\info{20,21}{Sebastian Kiehlmann}
\info{25}{Luca Latronico}
\info{6}{Alexander A. Lutovinov}
\info{20,21}{Nikos Mandarakas}
\info{14}{Frédéric Marin}
\info{26}{Andrea Marinucci}
\info{23}{Jon M.\ Miller}
\info{27}{Tsunefumi Mizuno}
\info{6}{Sergey V. Molkov}
\info{22}{Nicola Omodei}
\info{28}{Pierre-Olivier Petrucci}
\info{2}{Ajay Ratheesh}
\info{24}{Takeshi Sakanoi}
\info{6}{Andrei N. Semena}
\info{20,21}{Raphael Skalidis}
\info{2}{Paolo Soffitta}
\info{29}{Allyn F. Tennant}
\info{30}{Phillipp Thalhammer}
\info{31,32,33}{Francesco Tombesi}
\info{29}{Martin C. Weisskopf}
\info{30}{Joern Wilms}
\info{27}{Sixuan Zhang}
\info{34}{Iván Agudo}
\info{35,36}{Lucio A. Antonelli}
\info{37}{Matteo Bachetti}
\info{38,39}{Luca Baldini}
\info{29}{Wayne H. Baumgartner}
\info{38}{Ronaldo Bellazzini}
\info{29}{Stephen D. Bongiorno}
\info{25,40}{Raffaella Bonino}
\info{38}{Alessandro Brez}
\info{41,42,43}{Niccolò Bucciantini}
\info{38}{Simone Castellano}
\info{26}{Elisabetta Cavazzuti}
\info{32,36}{Stefano Ciprini}
\info{2}{Enrico Costa}
\info{2}{Alessandra De Rosa}
\info{2}{Ettore Del Monte}
\info{26}{Laura Di Gesu}
\info{2}{Alessandro Di Marco}
\info{26}{Immacolata Donnarumma}
\info{44,6}{Victor Doroshenko}
\info{29}{Steven R. Ehlert}
\info{13}{Teruaki Enoto}
\info{2}{Yuri Evangelista}
\info{2}{Riccardo Ferrazzoli}
\info{45}{Shuichi Gunji}
\info{46,\dag}{Kiyoshi Hayashida}
\info{47}{Jeremy Heyl}
\info{48}{Wataru Iwakiri}
\info{49,50}{Svetlana G. Jorstad}
\info{3}{Vladimir Karas}
\info{29}{Jeffery J. Kolodziejczak}
\info{2}{Fabio La Monaca}
\info{51}{Ioannis Liodakis}
\info{25}{Simone Maldera}
\info{38}{Alberto Manfreda}
\info{49}{Alan P. Marscher}
\info{52}{Herman L. Marshall}
\info{53}{Ikuyuki Mitsuishi}
\info{54}{Chi-Yung Ng}
\info{29}{Stephen L. O'Dell}
\info{25}{Chiara Oppedisano}
\info{35}{Alessandro Papitto}
\info{55}{George G. Pavlov}
\info{22}{Abel L. Peirson}
\info{36,35}{Matteo Perri}
\info{38}{Melissa Pesce-Rollins}
\info{37}{Maura Pilia}
\info{37}{Andrea Possenti}
\info{36}{Simonetta Puccetti}
\info{29}{Brian D. Ramsey}
\info{22}{Roger W. Romani}
\info{38}{Carmelo Sgrò}
\info{56}{Patrick Slane}
\info{38}{Gloria Spandre}
\info{15}{Toru Tamagawa}
\info{57}{Fabrizio Tavecchio}
\info{58}{Roberto Taverna}
\info{53}{Yuzuru Tawara}
\info{29}{Nicholas E. Thomas}
\info{37}{Alessio Trois}
\info{4,6}{Sergey Tsygankov}
\info{58,59}{Roberto Turolla}
\info{60}{Jacco Vink}
\info{59}{Kinwah Wu}
\info{61,2}{Fei Xie}
\infoe{59}{Silvia Zane}}\\[5ex]
\end{center}
\inst{1}{Physics Department and McDonnell Center for the Space Sciences, Washington University in St. Louis, St. Louis, MO 63130, USA}
\inst{2}{Istituto di Astrofisica e Planetologia Spaziali, 
Istituto Nazionale di Astrofisica,
00133 Roma, Italy}
\inst{3}{Astronomical Institute of the Czech Academy of Sciences, 
14100 Praha 4, Czech Republic}
\inst{4}{Department of Physics and Astronomy, 20014 University of Turku, Finland}
\inst{5}{Nordic Institute for Theoretical Physics (Nordita), Kungliga Tekniska Högskolan (KTH) Royal Institute of Technology and Stockholm University, 
SE-106 91 Stockholm, Sweden}
\inst{6}{Space Research Institute of the Russian Academy of Sciences, 
Moscow 117997, Russia}
\inst{7}{School of Mathematics, Statistics, and Physics, Newcastle University, Newcastle upon Tyne NE1 7RU, UK}
\inst{8}{Dipartimento di Matematica e Fisica, Universit\`a degli Studi Roma Tre, 
00146 Roma, Italy}
\inst{9}{California Institute of Technology, Pasadena, CA 91125, USA}
\inst{10}{University of Maryland, Baltimore County, Baltimore, MD 21250, USA}
\inst{11}{NASA Goddard Space Flight Center, Greenbelt, MD 20771, USA}
\inst{12}{Center for Research and Exploration in Space Science and Technology, NASA/GSFC, Greenbelt, MD 20771, USA}
\inst{13}{Rikagaku Kenkyūjyo (RIKEN) Cluster for Pioneering Research, 2-1 Hirosawa, Wako, Saitama 351-0198, Japan}
\inst{14}{Université de Strasbourg, Centre national de la recherche scientifique, Observatoire Astronomique de Strasbourg, Unité Mixte de Recherche 7550, 67000 Strasbourg, France}
\inst{15}{Astronomical Institute, Charles University, 
CZ-18000 Prague, Czech Republic}
\inst{16}{National Astronomical Observatories, Chinese Academy of Sciences, 
Beijing 100101, China}
\inst{17}{Leibniz-Institut für Sonnenphysik, D-79104 Freiburg, Germany}
\inst{18}{IRSOL Istituto Ricerche Solari Aldo e Cele Dacc\`o, Faculty of Informatics, Universit\`a della Svizzera italiana, Via Patocci 57, Locarno, Switzerland}
\inst{19}{Euler Institute, Faculty of Informatics, Universit\`a della Svizzera italiana, Via la Santa 1, 6962 Lugano, Switzerland}
\inst{20}{Institute of Astrophysics, Foundation for Research and Technology-Hellas, GR-71110 Heraklion, Greece}
\inst{21}{Department of Physics, Univ. of Crete, GR-70013 Heraklion, Greece}
\inst{22}{Department of Physics and Kavli Institute for Particle Astrophysics and Cosmology, Stanford University, Stanford, CA 94305, USA}
\inst{23}{Department of Astronomy, University of Michigan, 
Ann Arbor, MI 48109-1107, USA}
\inst{24}{School of Sciences, Tohoku University, Aoba-ku, 980-8578 Sendai, Japan}
\inst{25}{Istituto Nazionale di Fisica Nucleare, Sezione di Torino, 
10125 Torino, Italy}
\inst{26}{Agenzia Spaziale Italiana (ASI), 
00133 Roma, Italy}
\inst{27}{Hiroshima Astrophysical Science Center, Hiroshima University, 1-3-1 Kagamiyama, Higashi-Hiroshima, Hiroshima 739-8526, Japan}
\inst{28}{Institut de Planétologie et d’Astrophysique de Grenoble (IPAG),  Université Grenoble Alpes, Centre national de la recherche scientifique, 
38000 Grenoble, France}
\inst{29}{NASA Marshall Space Flight Center, Huntsville, AL 35812, USA}
\inst{30}{Dr. Karl Remeis-Observatory, Erlangen Centre for Astroparticle Physics, Universität Erlangen-Nürnberg, 
96049 Bamberg, Germany}
\inst{31}{Dipartimento di Fisica, Universit\`a degli Studi di Roma ``Tor Vergata'', 
00133 Roma, Italy}
\inst{32}{Istituto Nazionale di Fisica Nucleare, Sezione di Roma ``Tor Vergata'', 
00133 Roma, Italy}
\inst{33}{Department of Astronomy, University of Maryland, College Park, Maryland 20742, USA}
\inst{34}{Instituto de Astrofísica de Andalucía,
18008 Granada, Spain}
\inst{35}{INAF Osservatorio Astronomico di Roma, 
00078 Monte Porzio Catone, Roma, Italy}
\inst{36}{Space Science Data Center, Agenzia Spaziale Italiana, 
00133 Roma, Italy}
\inst{37}{INAF Osservatorio Astronomico di Cagliari, 
09047 Selargius, Cagliari, Italy}
\inst{38}{Istituto Nazionale di Fisica Nucleare, Sezione di Pisa, 
56127 Pisa, Italy}
\inst{39}{Dipartimento di Fisica, Università di Pisa, 
56127 Pisa, Italy}
\inst{40}{Dipartimento di Fisica, Università degli Studi di Torino, 
10125 Torino, Italy}
\inst{41}{INAF Osservatorio Astrofisico di Arcetri, 
50125 Firenze, Italy}
\inst{42}{Dipartimento di Fisica e Astronomia, Università degli Studi di Firenze, 
50019 Sesto Fiorentino, Firenze, Italy}
\inst{43}{Istituto Nazionale di Fisica Nucleare, Sezione di Firenze, 
50019 Sesto Fiorentino, Firenze, Italy}
\inst{44}{Institut f\"ur Astronomie und Astrophysik, Universität Tübingen, 
72076 T\"ubingen, Germany}
\inst{45}{Yamagata University, 1-4-12 Kojirakawa-machi, Yamagata-shi 990-8560, Japan}
\inst{46}{Osaka University, 1-1 Yamadaoka, Suita, Osaka 565-0871, Japan}
\inst{47}{University of British Columbia, Vancouver, BC V6T 1Z4, Canada}
\inst{48}{Department of Physics, Faculty of Science and Engineering, Chuo University, 1-13-27 Kasuga, Bunkyo-ku, Tokyo 112-8551, Japan}
\inst{49}{Institute for Astrophysical Research, Boston University, 
Boston, MA 02215, USA}
\inst{50}{Department of Astrophysics, St. Petersburg State University, 
Petrodvoretz, 198504 St. Petersburg, Russia}
\inst{51}{Finnish Centre for Astronomy with the European Southern Observatory (ESO), 20014 University of Turku, Finland}
\inst{52}{Kavli Institute for Astrophysics and Space Research, Massachusetts Institute of Technology, 
Cambridge, MA 02139, USA}
\inst{53}{Graduate School of Science, Division of Particle and Astrophysical Science, Nagoya University, Furo-cho, Chikusa-ku, Nagoya, Aichi 464-8602, Japan}
\inst{54}{Department of Physics, The University of Hong Kong, Pokfulam, Hong Kong}
\inst{55}{Department of Astronomy and Astrophysics, Pennsylvania State University, University Park, PA 16802, USA}
\inst{56}{Center for Astrophysics, Harvard \& Smithsonian, 
Cambridge, MA 02138, USA}
\inst{57}{INAF Osservatorio Astronomico di Brera, 
23807 Merate, Lecco, Italy}
\inst{58}{Dipartimento di Fisica e Astronomia, Università degli Studi di Padova, 
35131 Padova, Italy}
\inst{59}{Mullard Space Science Laboratory, University College London, Holmbury St Mary, Dorking, Surrey RH5 6NT, UK}
\inst{60}{Anton Pannekoek Institute for Astronomy, University of Amsterdam, 
1098 XH Amsterdam, The Netherlands}
\inst{61}{Guangxi Key Laboratory for Relativistic Astrophysics, School of Physical Science and Technology, Guangxi University, Nanning 530004, China}
\inst{\hbox{$\dag$}}{Deceased}
\inst{*}{Corresponding authors. Email: krawcz$@$wustl.edu, fabio.muleri$@$inaf.it, dovciak$@$astro.cas.cz, alexandra.veledina@utu.fi}
%
%
%
\begin{abstract}
{\bf A black hole  x-ray binary (XRB) system forms when gas is stripped from a normal star 
and accretes onto a black hole, which heats the gas sufficiently to emit x-rays. 
We report a polarimetric observation of the XRB \hbox{Cygnus X-1} using the Imaging x-ray Polarimetry Explorer. The electric field position angle
aligns with the outflowing jet, indicating that the jet is launched from the inner x-ray emitting region. The polarization degree is $\mathbf{4.01\pm0.20\%}$ at 2 to 8 kiloelectronvolts,
implying that the accretion disk is viewed closer to edge-on than the binary orbit.
The observations reveal that hot x-ray emitting plasma is spatially extended 
in a plane perpendicular to the jet axis, not parallel to the jet.}\\[0.5ex]
\end{abstract}
Cygnus X-1 (Cyg X-1, also catalogued as HD\,226868) is a bright and persistent x-ray source.
It is a binary system containing a 21.2$\pm 2.2$ solar-mass black hole in a 5.6 day 
orbit with a 40.6$^{+7.7}_{-7.1}$ solar-mass star and is located 
at a distance of 2.22$^{+0.18}_{-0.17}$\,kiloparsecs (kpc) \cite{2021Sci...371.1046M}.
Gas is stripped from the companion star; as it falls in the strong gravitational 
field of the black hole it forms an accretion disk that 
is heated to millions of kelvin. The hot incandescent gas emits x-rays. 
Previous analyses of the thermal x-ray flux, its energy spectrum, and the shape 
of the x-ray emission lines have indicated that the black hole in Cyg X-1 spins rapidly,
with a dimensionless spin parameter $a>0.92$ (close to the maximum possible 
value of 1) \cite{Gou_2014}. 
Cyg X-1 also produces two pencil-shaped outflows of magnetized plasma, 
called jets, that have been imaged in the radio band \cite{2001MNRAS.327.1273S}. 
It is thus classified as a microquasar, being analogous to much larger
radio-loud quasars, supermassive black holes with jets. 

Black hole x-ray binaries are observed in states of x-ray emission thought 
to correspond to different configurations of the accreting matter \cite{2007A&ARv..15....1D}.
In the soft state, the x-rays are dominated by thermal emission from the accretion disk.
The thermal emission is expected to be polarized because x-rays scatter off electrons
in the accretion disk \cite{1980ApJ...235..224C,2009ApJ...691..847L,2009ApJ...701.1175S}.
In the hard state, the x-ray emission is produced by single and multiple scatterings 
of photons (coming from the accretion disk or generated by electrons in the magnetic field) 
off electrons of hot coronal gas. Observations constrain the corona to be much hotter 
($k\,T_{\rm e} \sim$100\,keV, with $k$ being the Boltzmann constant 
and $T_{\rm e}$ the electron temperature) than the accretion disk 
($k\,T_{\rm d}\sim0.1$ keV).
The shape of the corona, and its location with respect to the accretion disk, 
are a matter of debate \cite{2007A&ARv..15....1D,2021SSRv..217...65B}, but 
can be constrained by x-ray polarimetry \cite{2010ApJ...712..908S}.
Reflection of x-rays emitted by the corona off the accretion disk 
produces an emission component that includes the iron K$\alpha$ 
fluorescence line at $\sim 6.4$ keV, which can constrain the velocity 
of the accretion disk gas orbiting the black hole and the time dilation 
close to the black hole.  The reflection component is also expected to be polarized \cite{1993MNRAS.260..663M,1996MNRAS.283..892P}. 

We report here on x-ray polarimetric observations of Cyg X-1 with the 
{Imaging X-ray Polarimetry Explorer} {(IXPE)} space telescope \cite{Weisskopf2022}.
Theoretical predictions of the Cyg X-1 polarization degree (in the 2--8 keV {IXPE} band)
are around 1\% or lower, depending on the emission state  \cite{2009ApJ...691..847L,2009ApJ...701.1175S,2010ApJ...712..908S,2022ApJ...934....4K}.
These expectations used an inclination angle 
(the angle between the black hole spin axis and the line of sight) of
$i=27\fdg5\pm0\fdg8$ inferred from optical observations of the binary system
\cite{2021Sci...371.1046M}. 
Earlier polarization observations with the OSO-8  gave polarization degree 2.44$\pm$1.07\% and
polarization angle (measured on
the plane of the sky from north to east) $-18\degr\pm13\degr$ at 2.6 keV 
\cite{Weisskopf1977pol,1980ApJ...238..710L} and a non-detection at higher energies  \cite{Chauvin2018}. 
IXPE observed Cyg X-1 from 2022 May 15 to 21 with an exposure time of $\sim$242\,ksec. 
The IXPE 2--8 keV observations were coordinated with  
simultaneous x-ray and gamma-ray observations covering 
the energy range 0.2--250 keV, including the 
Neutron Star Interior Composition Explorer Mission
(NICER, 0.2--12 keV), the Nuclear Spectroscopic 
Telescope Array (NuSTAR, 3--79 keV), 
the Swift X-ray Telescope (XRT, 0.2--10 keV),
the Astronomical Roentgen Telescope -- X-ray 
Concentrator (ART-XC, 4--30 keV) of the Spectrum-R\"ontgen-Gamma 
observatory (SRG), and the INTEGRAL Soft Gamma-Ray Imager (ISGRI, 30--80 keV) 
on the International Gamma-Ray Astrophysics Laboratory (INTEGRAL) space telescopes \cite{SM}.
Simultaneous optical observations were performed 
with the Double Image Polarimeter 2 (DIPol-2) 
polarimeter mounted on the Tohoku 60\,cm  telescope 
at the Haleakala Observatory and the
Robotic Polarimeter (RoboPol) at the 1.3 m telescope 
of the Skinakas observatory, Greece \cite{SM}.

During the observation campaign Cyg X-1 was highly variable over the entire 0.2--250\,keV 
energy range (Figure~\ref{f:lightcurves}). 
The source was in the hard x-ray state with a 
photon index of 1.6 (Table\,\ref{t:Spectrafitparams}) and a 0.2--250\,keV luminosity of 1.1\% of the Eddington 
luminosity (the luminosity at which the
radiation pressure on electrons equals the 
gravitational pull on the ions of the accreted material).
We detected linear polarization in the IXPE data with a $>20\sigma$ statistical confidence
(Figures \ref{f:datacube},\ref{f:stokes}). 
The 2--8 keV polarization degree is $4.01\pm0.20$\% at an electric field position angle of $-20\fdg7\pm1\fdg4$.
The polarization degree and angle are consistent with  the previous results of OSO-8 at 2.6 keV  \cite{1980ApJ...238..710L}.
The evidence for an increase of the polarization degree
with energy (Figures~\ref{f:datacube}, \ref{f:pd-fit}) is
significant on the 3.4\,$\sigma$ level \cite{SM}. 
We find a 2.4$\sigma$ indication that the polarization degree increases with the source flux 
(Figure \ref{fig:polfracang_brightdip}).

We find no evidence that the polarization properties depend on orbital phase of the binary system (Figure \ref{f:op}). This excludes the possibility that the observed 
x-ray polarization originates from the scattering of 
x-ray photons off the companion star or its wind, 
and shows that these effects do not measurably 
impact the polarization properties.

We calculated a suite of emission models and compared them to the observations \cite{SM}. 
We estimate that $>90$\% of the x-rays come from the inner $\sim$2,000\,km diameter 
region surrounding the $\sim$60\,km diameter black hole. 
We compare the orientation of the x-ray bright region (which we assume is 
determined by the x-ray polarization angle; Figure \ref{f:comp}) 
to the orientation of the billion-km-scale radio jet. 
We find that the x-ray polarization aligns with the radio 
jet to within $\sim$5\degr\ (Figure\,\ref{f:jet}).

We decomposed the broadband energy spectra 
observed simultaneously with IXPE, NICER, 
NuSTAR, and INTEGRAL into a multi-temperature 
black body component (thermal emission from the accretion disk),  a power-law component 
(from multiple Compton scattering events in the corona),
emission reflected off the accretion disk, and emission 
from more distant stationary plasma \cite{SM} (Figure\,\ref{f:eeuf01}).
We find that the coronal emission strongly
dominates in the IXPE energy band, contributing $\sim90$\% of the observed flux.
The accretion disk and reflected emission components contribute 
$<$1\% and $\sim$10\% of the emission, respectively. 
Therefore our polarization measurements are likely to be 
dominated by the coronal emission.

We analyzed the optical data in various wavelengths \cite{SM}, finding an intrinsic optical polarization degree 
of $\sim$1\% and polarization angle of $-24\degr$.
The uncertainties on these results are dominated by
systematic effects related to the choice of polarization
reference stars and are $\pm$0.1\% on the polarization degree
and $\pm13\degr$ on the polarization direction 
(Figures~\ref{fig:field_map}-\ref{fig:pol_parallax}, 
and Table~\ref{tab:dipol}).
The optical polarization direction is thought to indicate the orientation of the orbital axis projected 
onto the sky \cite{Kemp1979}.
We find it aligns with the x-ray polarization direction and the radio jet.

The alignment of the x-ray polarization with the radio jet indicates
that the inner x-ray emitting region is directly related to the radio jet.
If the x-ray polarization is perpendicular to the 
inner accretion disk plane, as favored in our models \cite{SM}, 
this implies that the inner accretion disk is 
perpendicular to the radio jet, at least on the plane of the sky.
This is consistent with the hypothesis that jets of microquasars (and, by extension, of quasars) 
are launched perpendicular to the inner accretion flow \cite{1984RvMP...56..255B}.  

Figure \ref{f:comp} compares our observed polarization with 
theoretical predictions made using models of the corona \cite{SM}.
We find that the only models that are consistent with the observations 
are those in which the coronal plasma is extended perpendicular 
to the jet axis, so probably parallel to the accretion disk. 
In these models, repeated scatterings 
in the plane of the corona polarize the x-rays perpendicular to that plane.
Two models are consistent with our observations: i) a hot 
corona sandwiching the accretion disk \cite{1991ApJ...380L..51H}, 
as predicted by numerical accretion disk simulations \cite{2021ApJ...922..270K} 
or ii) a composite accretion flow with a truncated cold, geometrically thin optically thick disk and an inner, geometrically thick 
but optically thin, laterally extended region of hot plasma, possibly produced by  
evaporation of the cold disk \cite{1994A&A...288..175M}.
If the jet is launched from the inner, magnetized region 
of the disk, the jet carrying away disk angular momentum 
could leave behind a radially extended hot and optically 
thin corona \cite{2010A&A...522A..38P}.

The polarization data rule out models in which the corona 
is a narrow plasma column or cone along the jet axis, 
or consists of two compact regions above and below the black hole. 
Our modeling of these scenarios accounts for the effect of the coronal emission reflecting 
off the accretion disk \cite{SM}. 
These models predict polarization degree well below the observed values. 
Models that produce high polarization degree predict polarization directions 
close to perpendicular to the jet axis, 
a decreasing polarization degree with energy, or both, 
and therefore disagree with the observations.

In our favored corona models, the high polarization
degree we observe requires that the x-ray bright region is seen at a higher inclination than the 
$\sim$27\degr\ inclination of the binary orbit.  
Sandwich corona models involving  the Compton scattering 
of disk photons with initial energies of $\sim 0.1$\,keV 
require inclinations exceeding 65\degr.
Truncated disk models invoking Compton scattering of disk or internally generated lower-energy  ($\sim$1--10~eV) 
synchrotron photons \cite{2013MNRAS.430.3196V}
can reproduce the observed polarization degree 
for inclinations $>$45\degr.
Compared to the models with disk photons, the larger number of scatterings required to energize lower-energy synchrotron 
photons to keV energies results in higher polarization 
degree in the IXPE energy band (Figure\,\ref{f:c1}) \cite{SM}.

Although the x-ray polarization, optical polarization, 
and radio jet approximately align in the plane of the sky,
the inclination of the x-ray bright region exceeds that of the binary orbit,
implying that the inner accretion flow is seen more edge on than the binary orbit.
Because the bodies of a stellar system typically orbit 
and spin around the same axis (as most planets in our solar system),
we consider potential explanations for the mismatch between 
the inner accretion disk inclination and the orbital inclination.

Stellar mass black holes are formed during supernovae. 
The supernova that occurred in Cyg X-1 might have
left the black hole with a misaligned spin. 
Gravitational effects could align the inner accretion flow angular 
momentum vector with the black hole spin vector \cite{1975ApJ...195L..65B}.
In this scenario, aligning the inner accretion disk angular momentum vector 
with the black hole spin vector would also align the radio jet 
produced by the inner accretion disk with the black hole spin vector.
Several, but not all, analyses of Cyg X-1 reflected emission spectra 
give inclinations consistent with our $i>45^{\circ}$ constraint \cite{2014ApJ...780...78T,2015ApJ...808....9P}.

An alternative explanation for the large inclination of the 
x-ray emitting region invokes the precession of the 
inner accretion flow with a period of much longer than the orbital period \cite{2006MNRAS.368.1025L}.
From our analysis of a 2--4 keV long-term x-ray light-curve 
we infer that the IXPE 
observations were performed close to the maximum 
inner disk inclination (Figure \ref{f:sop}) \cite{SM}. 
We tested the hypothesis that the inner flow 
precesses with an amplitude of $^>_{\sim}17\fdg5$ 
by performing an additional 86 ksec IXPE target 
of opportunity observation of Cyg X-1 from 2022 June 18 to 20,
33 days after the May observations, which corresponds to half of the current superorbital period \cite{SM}.
If this hypothesis is correct, we expected the polarization degree to 
drop from 4.01$\pm$0.20\% to $\ll$1\% due to the inclination changing 
from $i>45\degr$ in May to $i\lesssim 10\degr$ in June.
The observations showed the source in the same hard state with a 2--8 keV 
polarization degree and angle of $3.84\pm0.31$\% and  $-25\fdg7\pm2\fdg3$, 
respectively (Figure \ref{f:mj}) \cite{SM}.
The polarization degree thus stayed constant between the May and June observations 
within the statistical uncertainties of the observations.
We therefore disfavor the hypothesis that precession of the inner accretion flow
leads to the high polarization degree of the May observation. 
The combined May and June polarization degree and angle are  
$3.95\pm0.17$\% and $-22\fdg2\pm1\fdg2$, respectively (Figure \ref{f:mj}) \cite{SM}.

Several authors noted that optically thin synchrotron emission from the base of the jet
could contribute up to 5\% to the Cyg X-1 x-ray emission in the hard state \cite{2014MNRAS.438.2083R,2014MNRAS.442.3243Z}. Synchrotron emission from electrons 
gyrating around magnetic field lines is polarized perpendicular to those field lines. 
Our observation of the x-rays being polarized parallel to the jet axis would require 
synchrotron emission from a toroidal magnetic field wound around the jet axis. 
For this magnetic field geometry seen at an inclination of $27\fdg5$ the 
theoretical upper limit on the polarization degree of the synchrotron emission is 8\% \cite{2005MNRAS.360..869L}.
The jet thus contributes $<0.4\%$ of the observed polarization degree.  
Furthermore, if the almost constant jet emission was the main source of the observed 
polarization, we would expect that a rise in the x-ray flux from the inner accretion 
flow would lead to an overall smaller polarization degree -- contrary to the observed trend 
(Figure \ref{fig:polfracang_brightdip}).

To summarize: the polarized x-rays from the immediate surrounding of the black hole carry the imprint of the geometry of the emitting gas. We find that the x-ray bright plasma is extended perpendicular to the radio jet. The high observed polarization degree either implies a more edge-on viewing geometry than given by the optical data, or yet unknown physical effects responsible for production of the x-rays in accreting black hole systems. 
\clearpage
\renewcommand{\thefigure}{{\bf 1}}
\begin{figure}
\vspace*{-1cm}
\begin{center}
    \includegraphics[totalheight=14cm]{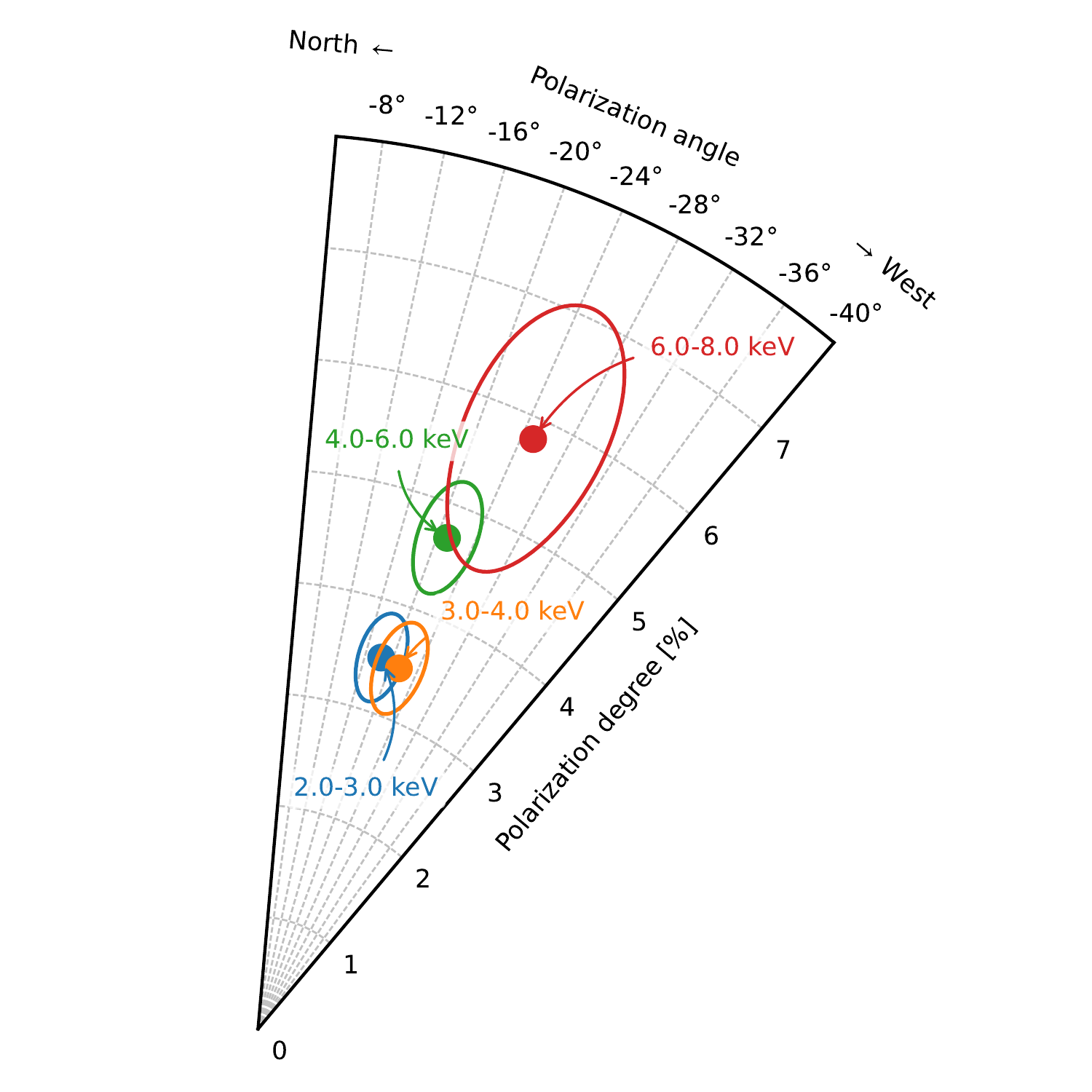}
    \end{center}
    \caption{\textbf{Energy-dependent x-ray polarization of Cyg X-1.}
    Polarization degree and polarization angle, derived from the IXPE observations, in four energy bands
    (labeled in different colours). 
    The ellipses denote the 68.3\% confidence regions.
     \label{f:datacube}}
\end{figure}

\renewcommand{\thefigure}{{\bf 2}}
\begin{figure}
\begin{center}
\vspace*{-1cm}
    \includegraphics[width=12cm]{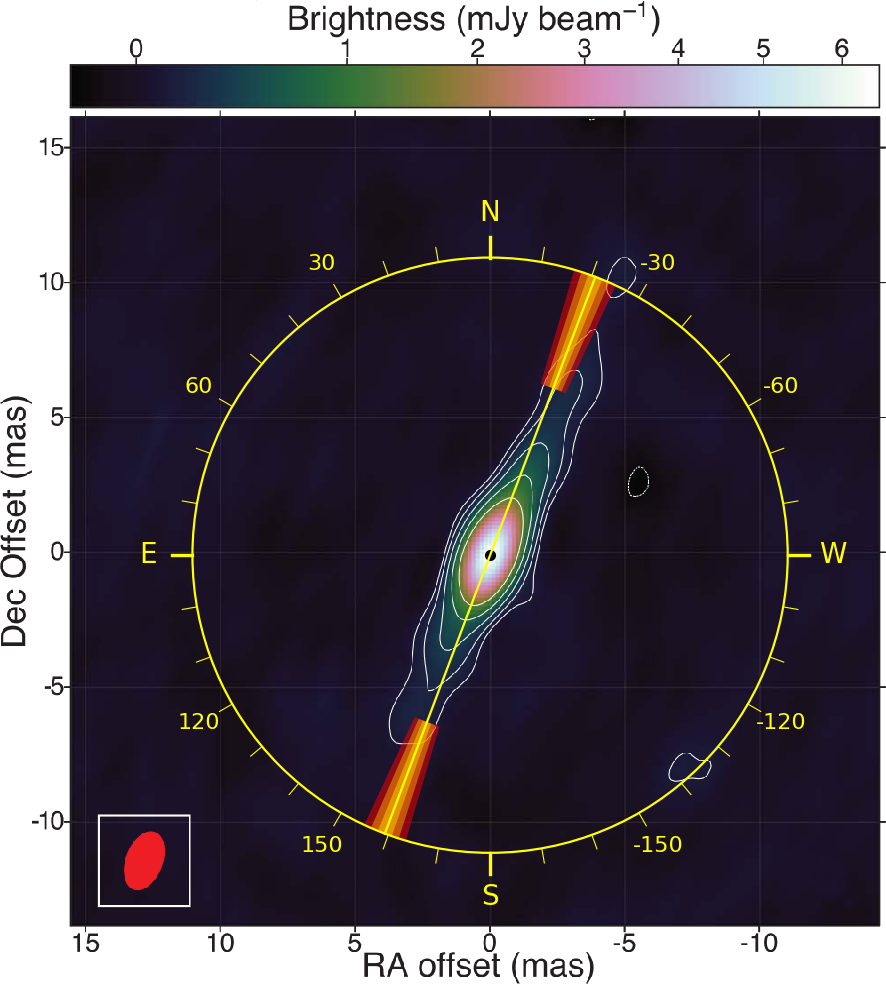}
\vspace*{-0.5cm}
    \end{center}
    \caption{\textbf{Comparison of the x-ray polarization direction with the radio jet \protect\cite{2021Sci...371.1046M}. }
    The 2--8 keV electric vector position angle is shown with the yellow line, and the one, two and three sigma 
    confidence intervals are given by the
    orange to red regions. 
    We infer (see text) that most x-rays are emitted by a $\sim$2,000\,km diameter region surrounding the $\sim$60 km diameter black hole, far smaller than the resolution of the radio image (indicated by the red ellipse).  
    The coordinate offsets in 
    right ascension (RA) and declination (Dec) are in the J2000 equinox
     in units of milli-arcseconds 
     (mas) with 1 arcsecond 
     being 1/3600$^{\rm th}$ 
     of a degree. 
     The color scale shows the radio 
     flux in milli-Jansky with 1 Jansky
     being 
     $10^{-26}$\,W\,m$^{-2}$\,Hz\,$^{-1}$.
     \label{f:jet}}
\end{figure}

\renewcommand{\thefigure}{{\bf 3}}
\begin{figure}
\begin{center}
\vspace*{-3cm}
\hspace*{-.8cm}
    \includegraphics[height=11cm]{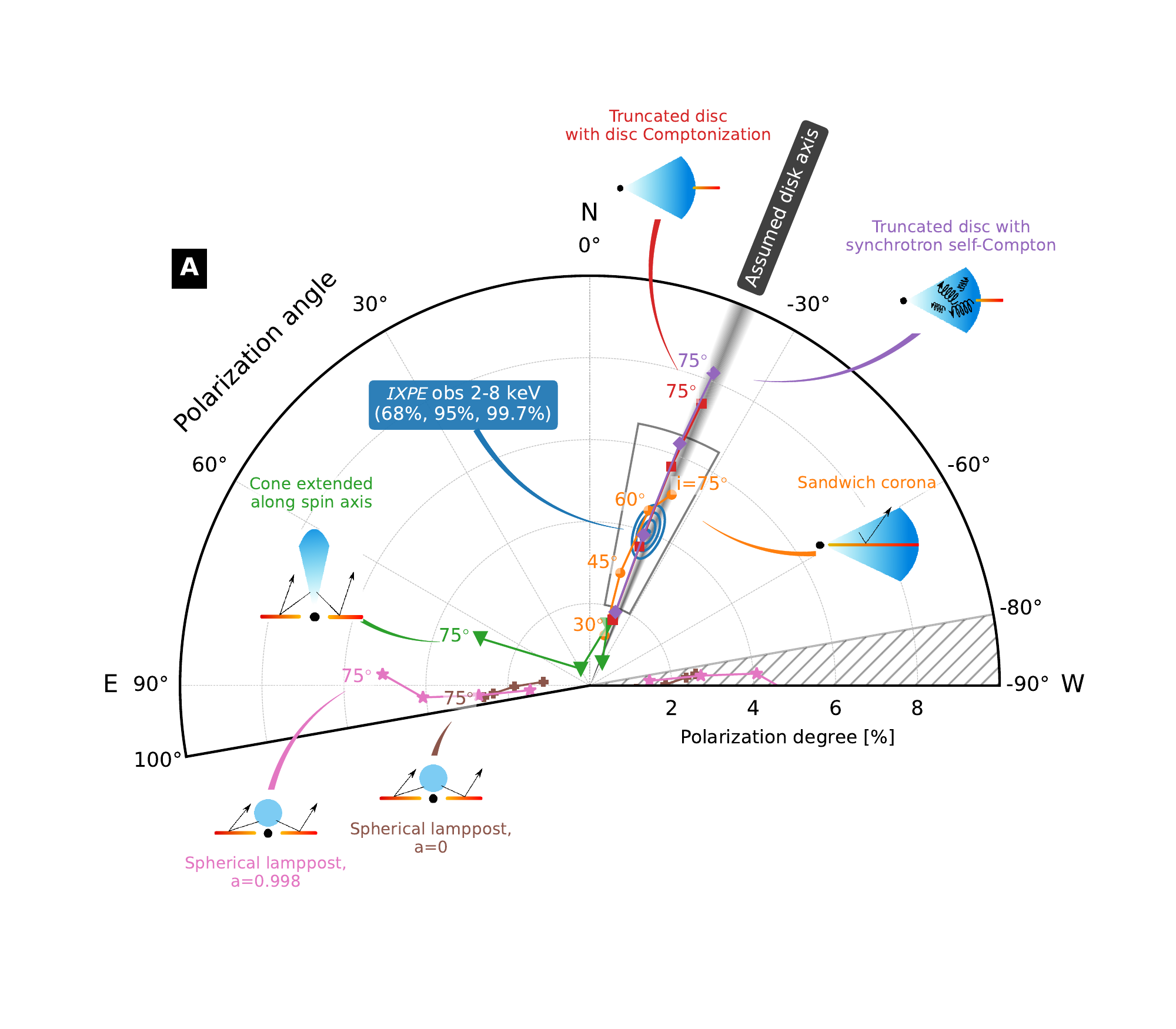}\hspace*{-1cm}
    \includegraphics[trim={5.8cm 0 0 0},clip,height=11cm]{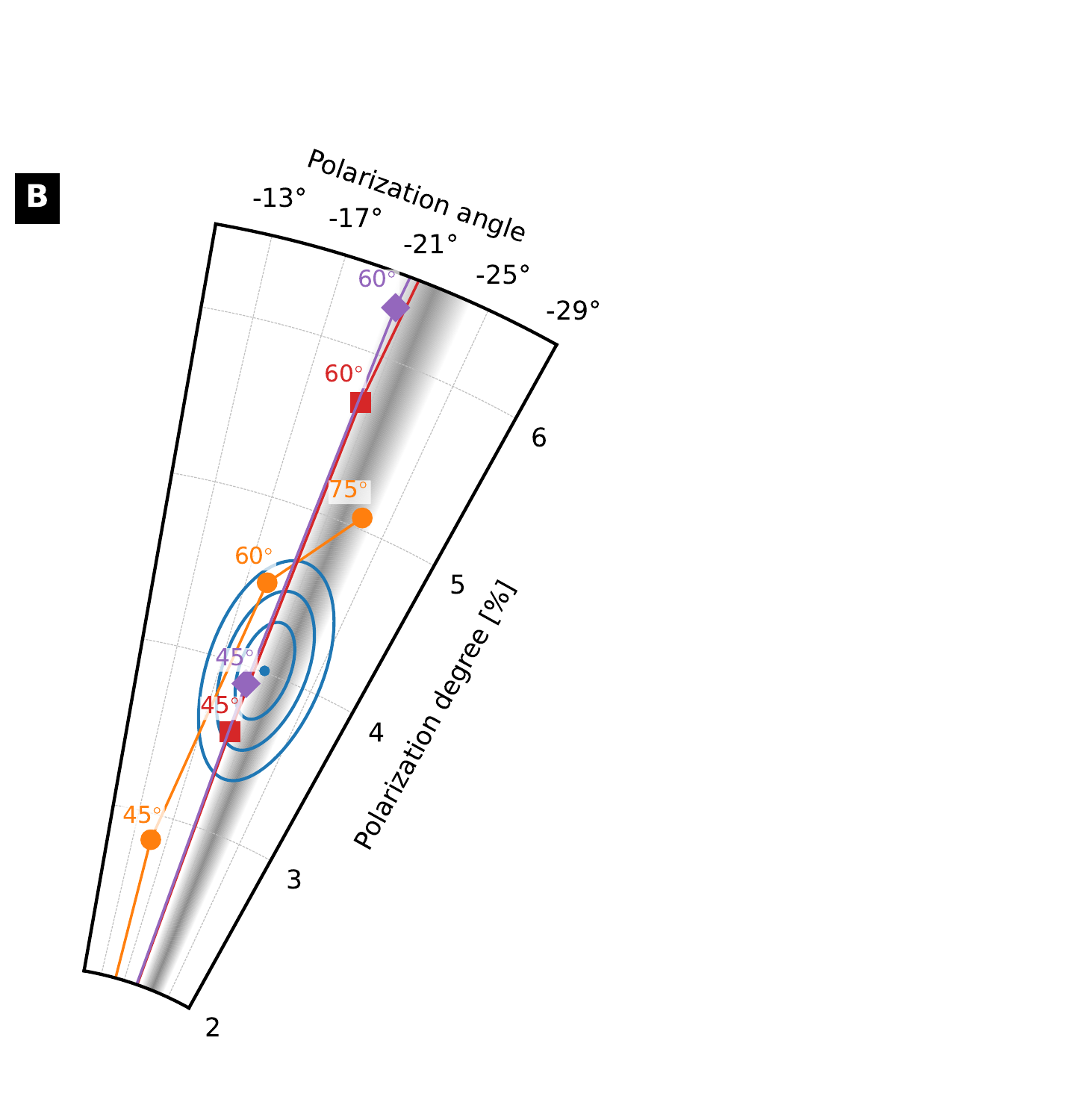}
\vspace*{-1cm}
    \end{center}
    \caption{
    \textbf{Comparison of the observed 2--8 keV polarization degree and angle with model predictions.} 
    (A) The blue dot shows the polarization 
    degree and angle, with the blue ellipses indicating the the 68\%, 95\% and 99.7\% 
    confidence levels (equivalent to 1$\sigma$, 
    2$\sigma$ and 3$\sigma$).
    Model predictions assume that the inner disk spin axis has position angle of $-22\degr$ 
    (consistent with the radio jet), 
    and that the inner disk angular momentum vector 
    points away from the observer  (as does the orbital angular momentum vector) \cite{2021Sci...371.1046M}. 
    The grey band shows the uncertainty 
    of the radio jet orientation; we adopt this as the uncertainty of the disk spin axis 
    in all models.
    Each colored line shows the results of 
    each chosen one corona geometry, 
    with symbols indicating different values
    as a function of the inner disk inclination $i$.
    Inset diagrams depict the assumed 
    black hole (black), corona (blue), and accretion disk (orange-red) configurations. 
    Black arrows indicate photon paths.
    Models with coronae extending parallel to the inner accretion disk can match the IXPE observations, but coronae located 
    or extending along the spin axis of the inner accretion disk cannot. 
    The position angles are shown from 
    $-80\degr$ to +100\degr\ (instead of $-90\degr$ to +90\degr) to show more clearly the models straddling the $\pm90\degr$ borders.
     (B) A zoom into the region around the measured value, marked with the grey box in panel A.
    }
\label{f:comp}
\end{figure}

\clearpage
\bibliographystyle{Science.bst}

\clearpage
{\bf Acknowledgements:}
The authors thank James Miller-Jones, Jerome Orosz and Andrzej Zdziarski for very helpful discussions of the optical constraints on the orbital inclination of Cyg X-1 and optical position angles.
We are grateful to three anonymous referees whose
excellent comments contributed to strengthening the paper. The authors thank Tom Maccarone for emphasizing that stellar wind absorption may 
modify the jet orientation measurement results.

The paper is based on the observations made with the IXPE mission, a joint US and Italian mission.  The US contribution to the IXPE mission is supported by the National Aeronautics and Space Administration (NASA) and led and managed by its Marshall Space Flight Center, with industry partner Ball Aerospace (contract NNM15AA18C). The Italian contribution to the IXPE mission is supported by the Italian Space Agency (Agenzia Spaziale Italiana, ASI) through contract ASI-OHBI-2017-12-I.0, agreements ASI-INAF-2017-12-H0 and ASI-INFN-2017.13-H0, and its Space Science Data Center (SSDC) with agreements ASI-INAF-2022-14-HH.0 and ASI-INFN 2021-43-HH.0, and by the Istituto Nazionale di Astrofisica (INAF) and the Istituto Nazionale di Fisica Nucleare (INFN) in Italy.

This research used data and software products or online services provided by the IXPE Team
    (Marshall Space Flight Center, the Space Science Data Center of the Italian Space Agency, the Istituto Nazionale di Astrofisica, and Istituto Nazionale di Fisica Nucleare),
    as well as the High-Energy Astrophysics Science Archive Research Center (HEASARC), at NASA Goddard Space Flight Center. We thank the NICER, NuSTAR, INTEGRAL,  { Swift}, and { SRG}/ART-XC teams and Science Operation Centers for their support of this observation campaign. DIPol-2 is a joint effort between University of Turku (Finland) and Leibniz Institut f\"{u}r Sonnenphysik (Germany).
We are grateful to the Institute for Astronomy, University of Hawaii, for allocating observing time for the DIPol-2 polarimeter, and the 
Skinakas observatory for performing the observations with the 
RoboPol polarimeter at their 1.3 m telescope.
\\[2ex]
{\bf Funding:} 
HK acknowledges NASA support under grants 80NSSC18K0264, 80NSSC22K1291, 80NSSC21K1817, and NNX16AC42G.
Fabio Muleri, JR, SB, SF, AR, Paolo Soffitta, EDM, EC, ADM, GM, YE, RF, FLM, Matteo Perri, and Alessio Trois were funded through the contract ASI-INAF-2017-12-H0. 
LB, Raffaella Bonino, Ronaldo Bellazzini, AB, LL, Simone Castellano, SM, Alberto Manfreda, CO, MPR, CS, and GS were funded by the ASI through the contracts ASI-INFN-2017.13-H0 and ASI-INFN 2021-43-HH.0.
MP was funded through the contract ASI-INAF-2022-14-HH.0.

IA acknowledges support from MICINN (Ministerio de Ciencia e Innovación) 
Severo Ochoa award 
for the IAA-CSIC (SEV-2017-0709) and through grants 
AYA2016-80889-P and PID2019-107847RB-C44.
MD, JS and Vladim{\'{\i}}r Karas acknowledge support from the GACR 
(Grantová agentura České republiky) project 21-06825X 
and the institutional support from the Astronomical Institute 
of the Czech Academy of Sciences (RVO:67985815). 
J.A.G.\ acknowledges support from NASA grant 80NSSC20K0540.
Jakub Podgorný acknowledges support from the Charles University project GA UK No. 174121, and from the Barrande Fellowship Programme of the Czech and French governments. AV, Juri Poutanen and SST thank the Russian Science Foundation grant 20-12-00364 and the Academy of Finland grants 333112, 347003, 349144, and 349906 for support. MN acknowledges support by NASA under award number 80GSFC21M0002.
TK is supported by JSPS KAKENHI Grant Number JP19K03902.
POP thanks for the support from the High Energy National Programme (PNHE) of Centre national de la recherche scientifique (CNRS) and from the French space agency (CNES) as well as from the
    Barrande Fellowship Programme of the Czech and French governments.
DB, SK, NM, RS acknowledge support from the European Research Council (ERC) under the European Unions Horizon 2020 research and innovation program under grant agreement No.~771282. Vadim Kravtsov thanks Vilho, Yrj\"o and Kalle V\"ais\"al\"a Foundation. PT and JW acknowledge funding from Bundesministerium f\"ur Wirtschaft and Klimaschutz under Deutsches Zentrum f\"ur
Luft- und Raumfahrt grant 50 OR 1909. 
AI acknowledges support from the Royal Society.
JH acknowledges the support of the Natural Sciences and 
Engineering Research Council of Canada (NSERC), 
funding reference number 5007110, and the Canadian Space Agency.
SGJ and APM are supported in part by the 
National Science Foundation grant AST-2108622, 
by the NASA Fermi Guest Investigator grant 80NSSC21K1917, 
and by the NASA Swift Guest Investigator grant 80NSSC22K0537.
C-YN is supported by a General Research Fund of the Hong Kong 
Government under grant number HKU 17305419. 
Patrick Slane acknowledges support from NASA Contract NAS8-03060.
\\[2ex]
{\bf Author contributions:} 
HK, Fabio Muleri, MD, AV, NRC, JS, AI, GM, JG, VL, and Juri Poutanen participated in the planning of the observation
campaign and the analysis and modeling of the data;
MN, TK, Jakub Podgorný, JR, and WZ contributed to the analysis or modeling of the data. AB, Vadim Kravtsov, SVB, MK, TS, DB, SK, NM, and RS
contributed to the optical polarimetric data.
JM and PD contributed the { Swift} results,
JW and PT  the INTEGRAL results, and AAL, SVM, 
and ANS the { SRG}/ART-XC results. 
SB, FC, NDL,  LL, AM, TM, NO, AR, POP, Paolo Soffitta, AFT, FT, MW, and SZ  contributed to the
discussion of the results. Frédéric Marin and SF served as internal referees. 
All the other authors contributed to the design, science case of the IXPE mission and in the planning of observations relevant to the present paper. All the authors provided input and comments on the manuscript.\\[2ex]
{\bf Competing interests:} The authors declare no conflicts of interest.\\[2ex]
{\bf Data and materials availability:}
The IXPE data are available at \url{https://heasarc.gsfc.nasa.gov/FTP/ixpe/data/obs/01/01002901/} and \url{https://heasarc.gsfc.nasa.gov/FTP/ixpe/data/obs/01/01250101/} for the May and June observations, respectively.
The NICER data are available at \url{https://heasarc.gsfc.nasa.gov/FTP/nicer/data/obs/2022_05/5100320101/}, \url{https://heasarc.gsfc.nasa.gov/FTP/nicer/data/obs/2022_05/5100320102/}, \url{https://heasarc.gsfc.nasa.gov/FTP/nicer/data/obs/2022_05/5100320103/}, \url{https://heasarc.gsfc.nasa.gov/FTP/nicer/data/obs/2022_05/5100320104/}, \url{https://heasarc.gsfc.nasa.gov/FTP/nicer/data/obs/2022_05/5100320105/}, \url{https://heasarc.gsfc.nasa.gov/FTP/nicer/data/obs/2022_05/5100320106/}, and \url{https://heasarc.gsfc.nasa.gov/FTP/nicer/data/obs/2022_05/5100320107/}.
The NuSTAR data are available at 
\url{https://heasarc.gsfc.nasa.gov/FTP/nustar/data/obs/07/3/30702017002},
\url{https://heasarc.gsfc.nasa.gov/FTP/nustar/data/obs/07/3/30702017004}, and
\url{https://heasarc.gsfc.nasa.gov/FTP/nustar/data/obs/07/3/30702017006}.
The SWIFT XRT data of Cyg X-1 are available at \url{https://heasarc.gsfc.nasa.gov/FTP/swift/data/obs/2022_05/00034310009/xrt/}, \url{https://heasarc.gsfc.nasa.gov/FTP/swift/data/obs/2022_05/00034310010/xrt/}, \url{https://heasarc.gsfc.nasa.gov/FTP/swift/data/obs/2022_05/00034310011/xrt/}, \url{https://heasarc.gsfc.nasa.gov/FTP/swift/data/obs/2022_05/00034310012/xrt/}, \url{https://heasarc.gsfc.nasa.gov/FTP/swift/data/obs/2022_05/00034310013/xrt/}, and \url{https://heasarc.gsfc.nasa.gov/FTP/swift/data/obs/2022_05/00034310014/xrt/}. 
The extracted INTEGRAL ISGRI data are available at Zenodo: \url{https://zenodo.org/record/7140274#.Yzsg0OzMI-Q} \cite{isgri}.
The SRG ART-XC data are available through the ftp server \url{ftp://hea.iki.rssi.ru/public/SRG/ART-XC/data/Cygnus_X-1/}.
The MAXI light curves are available through
\url{http://maxi.riken.jp/top/lc.html}.
The raw DIPol-2 data are available at Zenodo: \url{https://zenodo.org/record/7108247} \cite{Kravtsov22}.
The raw RoboPol data are available at Zenodo: \url{https://zenodo.org/record/7127802} \cite{robopol}.
%
%
The \textsc{kerrC} x-ray model \cite{2022ApJ...934....4K} is available at
\url{https://gitlab.com/krawcz/kerrc-x-ray-fitting-code.git}.
The \textsc{MONK} x-ray model \cite{2019ApJ...875..148Z} is available at \url{https://projects.asu.cas.cz/zhang/monk}.
Models of polarized emission in the truncated disk geometry \cite{VeledinaPoutanen22}
are available at Zenodo: \url{https://zenodo.org/record/7116125} .

Our derived x-ray polarization measurements are listed in Tables S1 and S2, and the optical polarization in Table S4. The numerical results of our model fitting are listed in Table S5. 
\\[2ex]
{\bf Supplementary Materials:}
Materials and Methods, Figures S1 to S12, Tables S1 to S5,
References (37-79).
\clearpage

\renewcommand{\thepage}{S\arabic{page}} 
\setcounter{page}{1}

{\Large Supplementary Materials for} \\[4ex]
\textbf{\large Polarized x-rays constrain the disk-jet geometry in the black hole x-ray binary Cygnus X-1}\\[4ex]
{Henric Krawczynski$^*$, Fabio Muleri$^{**}$, Michal Dovčiak$^{\dag}$, Alexandra Veledina$^{\ddag}$, \textit{et al.}}\\[4ex]
{Corresponding authors. 
Emails: $^*$krawcz@wustl.edu, 
$^{**}$fabio.muleri@inaf.it, 
$^{\dag}$dovciak@astro.cas.cz,
$^{\ddag}$alexandra.veledina$@$utu.fi}
\hspace*{0.5cm}\\[4ex]
\thispagestyle{empty}
\noindent {\bf The PDF file includes:}\\
Materials and Methods\\
Figures S1 to S12\\
Tables S1 to S5

\clearpage

\section*{Materials and Methods}
\label{sec:methods}
\subsubsection*{Data Sets and Analysis Methods} 
IXPE observed Cyg X-1 from 2022 May 15 to 21 for 242\,ksec.
Following the results from the May IXPE observation campaign, 
we performed an additional 86\,ksec target of opportunity 
observation of Cyg X-1 from 2022 June 18 to 20. 

The spectral fitting of the IXPE data uses the level 2 IXPE data
and the software tools \xspec\  \cite{1996ASPC..101...17A} 
and {\tt Sherpa} \cite{2001SPIE.4477...76F,2007ASPC..376..543D,SciPyProceedings_51,brefsdal-proc-scipy-2011}. 
The model-independent Stokes
parameter analysis \cite{2015APh....68...45K} of the 
IXPE polarization data was performed
with the {\tt ixpeobssim} software \cite{Baldini2022}. 
The {\tt ixpeobssim}\textbackslash{\tt xpbin} command \cite{2015APh....68...45K,Baldini2022} is used to extract
Stokes parameters and the polarization degree and angle from the Level 2 data. The confidence regions for the polarization measurements were calculated using standard methods 
\cite{Weisskopf2010, Muleri2022}.
The results were cross-checked by fitting 
the Stokes $I$, $Q$ and $U$ data
with \xspec\ using the 
response matrices from  the 
High Energy Astrophysics Science Archive Research Center (HEASARC)
data archive \cite{2017ApJ...838...72S}. 
Source and background data were selected based on the reconstructed arrival
direction in celestial coordinates. The source events were selected with a circular 
region of $\sim$80 arcsec radius; background events were selected with a concentric annulus 
of inner and outer radii of $\sim$150 and $\sim$310 arcsec, respectively.
We use the additive property of the Stokes parameters to subtract the background.
The signal exceeds the background by $>$70 times over the entire energy range of the 
polarization measurements.

The NuSTAR spacecraft 
\cite{2013ApJ...770..103H}
acquired a total of ~42\,ksec of data between 2022 May 18 and May 21. The NuSTAR data were processed with the {\tt NuSTARDAS} software (version 1.9.7) of the {\tt HEAsoft} package (version 6.30.1)\cite{heasoft}.

NICER \cite{2012SPIE.8443E..13G} acquired a total of ~87\,ksec of data between May 15 and May 21, 2022. The NICER data were processed with the  {\tt NICERDAS} software (version 9.0) of the {\tt HEASoft} package.).

{ Swift} observed Cyg X-1 daily between May 15 and May 20, 2022 for a total of $\sim$54\;ksec, with the XRT instrument operating in Windowed Timing (WT) mode. The observations were processed using the tools in {\tt HEASoft}. The initial event cleaning was performed using \texttt{XRTPIPELINE}, the spectra and light curves were extracted using \texttt{XSELECT}, and ancillary response files (ARF) were generated using \texttt{XRTMKARF}.

The Mikhail Pavlinsky ART-XC telescope \cite{Pavlinsky21} on board the {SRG} observatory \cite{Sunyaev21} carried out two observations of Cyg~X-1 on 2022 May 15 to 16 and 18 to 19, simultaneous with IXPE, with 86 and 85 ks exposures, respectively. 
ART-XC data were processed with the analysis software \texttt{ARTPRODUCTS} v0.9 with the CALDB (calibration data base) version 20200401. 

INTEGRAL observed Cyg X-1 between 2022 May 15 and May 20
with a total exposure time of $\sim$196\,ksec. INTEGRAL/ISGRI light curves and energy spectra were extracted using version 11.2 of the \textsc{Off-line Scientific Analysis} (OSA) software \cite{integral}.

We used the Cyg X-1 observations with the 
Monitor of All-sky X-ray Image (MAXI) \cite{2009PASJ...61..999M} 
to extract a long-term 2--4 keV light curve (Figure \ref{f:sop}). 
Figure \ref{f:lightcurves} shows the IXPE, 
NICER, NuSTAR, {Swift}/XRT, {SRG}/ART-XC, and INTEGRAL light curves. 

As mentioned in the main article, we used IXPE to test 
the hypothesis that the high polarization fraction of the May 15-21 IXPE observations was caused by the superorbital (i.e. with a period exceeding the orbital period) precession of the inner accretion flow \cite{1991MNRAS.249...25W,2012MNRAS.420.1575K}.
Cyg X-1 exhibits superorbital flux modulations
that are stable over periods of years \cite{2006MNRAS.368.1025L,2008MNRAS.389.1427P}.

Figure \ref{f:sop} shows the Cyg X-1 2--4 keV flux between December 17, 2020 and August 9, 2022. The blue dashed lines show the dates of the
fitted superorbital flux minima.
The green solid lines indicate the time of the first 
(May 15--21) and second (June 18--20) IXPE observation campaigns,
close to the time of a superorbital flux minimum 
(first observation) and maximum (second observation).
If the inner accretion flow indeed precesses, the superorbital
flux minimum should correspond to inclination and 
polarization degree maxima, and the  superorbital
flux maximum should correspond to inclination and 
polarization degree minima.
As described in the main text, the IXPE observations did not
show the drastic change of the polarization degree 
predicted by the precession hypothesis.
\subsubsection*{IXPE Polarization Results \label{sm:datacube}}
Figure \ref{f:stokes} shows the IXPE polarization signal from the May 15 to May 21, 2022
observations in terms of the normalized Stokes parameters $Q/I$ and $U/I$, 
giving the polarized beam intensity along the north-south ($Q/I>0$) and east-west ($Q/I<0$) 
directions as well as along the northeast--southwest 
($U/I>0$) and northwest--southeast ($U/I<0$) directions.
Tables~\ref{tab:Stokes} and \ref{tab:pol}
give the results of both analyses in terms
of the Stokes parameters, and polarization degree and angle, respectively.
The consistency of the radio-jet -- x-ray polarization alignment is limited 
by the precision of the radio results. 
Different studies have found $-26\degr$ \cite{2021Sci...371.1046M}, or 
$-21\degr$ to $-24\degr$ in 3 epochs, but $-17\degr$ 
for the inner jet in another epoch \cite{2001MNRAS.327.1273S}. 
The variability of the results could be explained by the phase dependent absorption 
of the radio emission by the stellar wind \cite{2021Sci...371.1046M}. 

The target of opportunity observations of Cyg X-1 from June 18 to 20, 2022 
showed the source still in the hard state. We measure a polarization degree and angle of 
3.84$\pm$0.31\% and $-25\fdg7 \pm 2\fdg3$, respectively, for this data set.
We present the results from the May and June observations as well as the results from the
cumulative data set in Figure \ref{f:mj}. The results are consistent with time 
independent polarization degree and polarization angle.
The polarization degree and direction of the cumulative data set are
3.95$\pm$0.17\% and $-22\fdg2 \pm 1\fdg2$, respectively.

In the following we limit the analysis to the data acquired in May to avoid merging data taken a 
month apart. The polarization degree increases with 
energy from 3.5$\pm$0.2\% in the energy band 2--5 keV to 
5.3$\pm$0.5\% in the energy band 5--8 keV \cite{SM}.
Fitting a model of constant polarization is rejected 
at the 99.93\% confidence level.
The polarization degree (PD) increase with energy 
is better matched by a linear model $PD\,=\,A + B \times (E/{\rm keV}-1)$ with $A = (2.9 \pm 0.4)\%$ and $B = (0.58 \pm 0.15)\%$ (Figure~\ref{f:pd-fit} A).
On theoretical grounds, we expect that the x-ray 
emission around the  Fe K$\alpha$ line energy of 
6.4~keV exhibits a reduced polarization degree. 
We find however, that the dips of the polarization degree at 
4.5--5 and 6--6.5\,keV are not statistically significant.
The fit of a linear function has a $\chi^2$ of 4.04 for 9 degrees of 
freedom and a chance probability of larger $\chi^2$-values of 90.9\%.
Moreover, based on the constraints on the equivalent width 
of the fluorescent Fe K$\alpha$-line from the 
spectral analysis of the NICER and NuSTAR data, 
we find that the maximum possible Fe K$\alpha$ depolarization 
is much smaller than the observed dips.
A fit of the polarization angle as a function of energy with 
a constant function gives a statistically acceptable fit with a
chance probability for larger $\chi^2$-values of 57.5\% 
(Figure~\ref{f:pd-fit} B).

The light curves in Figure~\ref{f:lightcurves} show  that the Cyg X-1 IXPE 
count rates varied between 20 and 60 count s$^{-1}$.
We investigated the flux dependence of the polarization properties by analyzing three count-rate selected data sets.
The average fluxes of those data sets are 3.5, 3.9, and 4.5 times 
$10^{-9}$ erg cm$^{-2}$ s$^{-1}$.
The polarization degree increase with the flux from 
$3.63\pm0.30$\% to $3.87\pm0.34$\% 
to $5.03\pm0.41$\% 
(Figure \ref{fig:polfracang_brightdip}).
The overall trend is statistically significant 
at the 98.3\% confidence level.

Figure~\ref{f:op} shows that the polarization properties 
(Stokes $Q/I$ and $U/I$) do not depend on the orbital 
phase of the binary. Fitting the polarization along 
the orbit with a constant  provides an acceptable null 
hypothesis probability. Data are summed 
between 2 and 8 keV. The assumed period is 5.599829~days, with $T_0$ at MJD 52872.288 \cite{2008ApJ...678.1237G}.
\subsubsection*{IXPE, NICER, NuSTAR, and INTEGRAL  energy spectra}

We used the \textsc{XSPEC} package for fitting a simple 
model to the broadband Stokes $I$ spectrum provided 
by NICER, IXPE, NuSTAR, and INTEGRAL 
and the Stokes $Q$ and $U$ spectra provided 
only by IXPE. 
We use the data from the first NuSTAR observation 
and the simultaneously acquired NICER data,  
to eliminate differences due to spectral variability. 
We use the entire IXPE and INTEGRAL observations to maximize 
the signal-to-noise ratio. 
We fit the two NuSTAR Focal Plane Modules (FPMs) 
and the three IXPE detector inits separately in the fit. 
For the Stokes {\it I} spectrum, we employ the \textsc{XSPEC} fitting models
\begin{equation}
\textsc{mbpo} * \textsc{tbabs} * ( \textsc{diskbb} + \textsc{xillverCp} + \textsc{relxillCp} + \textsc{nthComp} ).
\label{eqn:specpol}
\end{equation}
Here \textsc{diskbb} represents thermal disk emission and \textsc{nthComp} represents Compton scattered emission observed directly from the corona. The \textsc{relxillCp} component represents coronal x-rays that are reflected from the inner accretion disk and distorted by relativistic effects. 
We assume that the flux irradiating the disk 
decreases with increasing radial distance proportional to $r^{-3}$.
The \textsc{xillverCp} component represents coronal x-rays that are reflected from the outer disk and the companion star and not subject to strong relativistic effects. \textsc{tbabs} accounts for line-of-sight absorption by the interstellar medium.

The model \textsc{mbpo} is included to account for cross-calibration discrepancies we encountered between the four observatories. It multiplies the model spectrum by a broken power law, \textsc{mbpo}$(E) = N (E/E_{\rm br})^{\Delta\Gamma}$, 
where $E$ is the energy of the photon and $N$ is a normalization constant 
giving the ratio of the detection 
areas of the satellites 
at the energy $E_{\rm br}$ at which the
power law index of the model 
changes from the value 
$\Delta\Gamma_1$ to $\Delta\Gamma_2$.
For NICER, we fix the power-law indices to zero and the normalization to unity. For each NuSTAR FPM and INTEGRAL, we tie $\Delta\Gamma_2=\Delta\Gamma_1$ (i.e. employing only a single power law) but leave $\Delta\Gamma_1$ and $N$ as free parameters of the fit.
For the IXPE detector units, we leave all \textsc{mbpo} parameters free. 
We also include a $0.5\%$ systematic uncertainty to further account for cross-calibration discrepancies. Finally, the NuSTAR FPM~A disagrees with the FPM~B and NICER in the 3--4 keV band, and IXPE detector unit \#3 disagrees with all other instruments (even with the use of \textsc{mbpo}) in the $>5$ keV energy range, and so we ignore these ranges in our model fitting.

We first jointly fit the model to the NICER, NuSTAR and INTEGRAL data, then add IXPE Stokes $I$ to fit the model before finally adding IXPE Stokes $Q$ and $U$. At each stage, the best-fit parameters  change by less than their uncertainties.
We tie the seed photon temperature of the \textsc{nthComp} component
(parameter $k T_{\rm bb}$) to the temperature of the inner edge 
of the accretion disk 
(parameter $k\,T_{\rm d}$ of the \textsc{diskbb} model).
We tie the \textsc{relxillCp} photon index to that of the \textsc{nthComp} component, but are unable to do this for the seed photon temperature as this hardwired to $0.05$\,keV in the \textsc{relxillCp} grid. We initially forced the \textsc{relxillCp} and \textsc{nthComp} components to have the same coronal electron temperature 
$k\,T_{\rm e}$, but found that the fit improved dramatically ($\gg 5~\sigma$ according to an F-test) after relaxing this assumption. 
The discrepancy between the corona temperature seen by the observer (\textsc{nthComp} temperature of 94\,keV) and by the disc (\textsc{relxillCp} temperature of 140\,keV) may be due to general relativistic effects 
(redshifting the emission seen by the observer),
and due to the different viewing angles of the corona.
We calculate $90\%$ confidence level uncertainties on the fitting results 
with a Markov Chain Monte Carlo simulation that uses the Goodman-Were 
algorithm with a total length of 307,200 steps spread over 256 walkers 
following an initial burn-in period of 19,968 steps. The best-fit spectral parameters are listed in Table \ref{t:Spectrafitparams}.

Figure~\ref{f:eeuf01}a shows the best-fit Stokes $I$ model
and the data unfolded around that model, as well as the
contributions from the different model components.
The \textsc{diskbb}, \textsc{xillverCp} and \textsc{relxillCp} components contribute respectively 0.6\%, 0.5\% and 10.0\% of the flux. The fractional contribution of each model component is consistent whether we consider only NICER, NuSTAR and INTEGRAL or also include IXPE. Because the direct coronal flux dominates the 2--8 keV flux, it must also dominate the polarization. For instance, the relativistic reflection component would need to be $\sim 40\%$ polarized to achieve the observed overall polarization of $\sim 4\%$. 
However, the reflected emission exhibits most likely much smaller polarization degree \cite{1993MNRAS.260..663M,1996MNRAS.283..892P,2011ApJ...731...75D,2022MNRAS.510.4723P} 
(see also Figures \ref{f:c1} and \ref{f:c2}).

As a simple toy model, we therefore assign a constant (independent of energy) polarization degree and angle to the \textsc{nthComp} component (the model \textsc{polconst}) and assume that the other components are unpolarized. Fig.\,\ref{f:eeuf01}c shows the resulting fit to IXPE Stokes {\it Q} and {\it U}. 
We find a reduced $\chi^2$ of $\chi^2/({\rm degrees\,of\,freedom)}=2575.72/2466$. 
Panel Fig\,\ref{f:eeuf01}d shows the contributions 
from each energy channel to $\chi$, we find that there are no structured residuals. The best-fit polarization degree and angle of the corona from this simple model are respectively $3.63\pm 0.26\%$ and $ -20\fdg5\pm 2\fdg1$ ($90\%$ confidence).

\subsubsection*{Model constraints on the inclination of the inner accretion disk}
We studied the energy spectra and polarization properties of different corona shapes and properties with the 
raytracing codes 
\textsc{kerrC} \cite{2022ApJ...934....4K}, 
\textsc{Monk} \cite{2019ApJ...875..148Z}, 
and with an iterative radiation transport solver   
\cite{1996ApJ...470..249P,VeledinaPoutanen22}.
We present simulation results that match the IXPE, 
NICER, and NuSTAR energy spectra qualitatively, 
and the predicted polarization properties.

The Cyg X-1 binary system spins clockwise \cite{2021Sci...371.1046M}; we therefore plot position angles assuming that the inner disk 
and the black hole also spin clockwise. 
This assumption impacts the sign of the
predicted polarization angles. We assume furthermore
that the inner disk and black hole spin axes are aligned
and are at 0\degr\ position angle. 
The position angles shown in Figure \ref{f:comp} were 
obtained by  subtracting $22\degr$ from the position 
angles in the models.

We used the general relativistic ray tracing codes \textsc{kerrC}
to evaluate the polarization that cone-shaped coronae centered 
on the black hole spin axes and wedge-shaped coronae sandwiching the accretion disk can produce. 
The code assumes a standard geometrically thin, optically thick accretion disk
extending from the innermost stable circular orbit to 100 gravitational radii $r_{\rm g}=G\,M/c^2$
with $G$ being the gravitational constant, $M$ the black hole mass, and $c$ the speed of light.
The code uses Monte Carlo methods to simulate the polarized emission of the accretion disk photons
assuming Novikov-Thorne temperature profiles,  the geodesic propagation of the x-rays 
including the general relativistic polarization direction 
evolution, the polarization-changing Compton scattering of the photons in the corona, 
and the reflection of the photons off the accretion disk adopting the {\tt XILLVER} reflection model
for the reflected intensity \cite{2011ApJ...731..131G,2013ApJ...768..146G,2014ApJ...782...76G}, and an analytical solution for the reflected polarization \cite{1960ratr.book.....C}.
In both cases, we chose corona parameters which  maximize the predicted polarization degree, i.e., cone-shaped coronae close to the accretion disk, and thin wedge-shaped coronae 
with a half opening angle of 10\degr. The model parameters are given in Table \ref{t:kerrc}. For all models, we assume that the black hole spin
vector and the inner disk spin vector are aligned.
The sandwich and cone corona models (as well as the extended lamppost corona model discussed below) are phenomenological - the coronal temperatures are not derived self-consistently. Coronae could cool radiatively, to the point that the predicted energy spectra are softer than the observed ones \cite{Malzac05,Poutanen18}. Processes that heat and cool the coronal plasma are debated, as are their relative contributions \cite{2021LRCA....7....1N,2021ApJ...922..270K,2022arXiv220302856S}.

We also used the ray tracing code \textsc{Monk}, which is similar 
to \textsc{kerrC} but implements the
simulation of an extended lamppost corona. The lamppost corona is centered on the spin axis of the accretion disk
at a radial coordinate of $r\,=\,$10$\,r_{\rm g}$ and has a radius of 8 $r_{\rm g}$, an electron temperature of 100 keV, and Thomson optical depth of 1 (defined as $n_e \sigma_{\rm T} R_c$, 
where $n_e$ is the electron density of the corona, 
$\sigma_{\rm T}$ is the Thomson cross section,
and $R_c$ is the radius of the corona). Simulations were performed for both Schwarzschild ($a=0$) and Kerr ($a=0.998$) black holes, with mass accretion rate of $4.71\times10^{17}$ and $2.64\times 10^{18}~\rm g~s^{-1}$, respectively. 
For the \textsc{Monk} simulations, we first calculated the
Stokes parameters generated by the direct emission and then added those
of the reflected emission. The reflected emission was normalized to reproduce  the reflected emission fraction from the analysis of the NICER,
IXPE, NuSTAR, and INTEGRAL energy spectra.
We compared the \textsc{Monk} results before and after accounting for the
reflected emission. The reflected emission 
lowers the total polarization degree by $\sim$20\% 
(e.g.\, 
a polarization degree of 3\% before accounting for 
reflection becomes 2.5\% after accounting for the impact
of reflection) as the different
polarization directions of the direct and reflected 
emission components
lead to the partial cancellation of the different polarizations.

We studied the polarization of the truncated disk/inner hot flow scenario with the iterative radiation transport solver mentioned
above.  The code treats Compton scattering of polarized 
radiation in a plane-parallel geometry in flat space. 
It uses exact Compton scattering redistribution matrices for isotropic electrons \cite{1993A&A...275..325N} and solves the polarized radiation 
transfer equations using an expansion of the intensities in 
scattering orders. 
We do not include reflection off the
cold disk \cite{1996MNRAS.283..892P}
to avoid uncertainties related to the 
properties of the reflecting plasma.
The code simulates a plane parallel slab, 
using a prescription to inject seed photons that mimics 
the truncated disk scenario with the hot flow height-to-radius ratio of 1. 
The electron temperature is assumed to be $kT_{\rm e}=100$~keV, 
the seed blackbody temperature $kT_{\rm bb}=0.1$~keV 
and the Thomson optical depth $\tau_{\rm T}=1.0$
\cite{Gierlinski1997,2002ApJ...578..357Z}.
Analytical prescriptions are used to account for the impact of special and general relativistic effects on the observed
polarization degree and angle \cite{Loktev2022} in the Schwarzschild metrics.

Figures \ref{f:c1} and \ref{f:c2} summarize the polarization predictions.
Figure \ref{f:c1} shows the simulation results for models with coronae  extending parallel to the accretion disk.
The sandwich corona simulated with \textsc{kerrC} 
generates sufficiently large 
polarization degree for $i\gtrsim60\degr$. 
The polarization direction aligns within a few degrees 
with the inner disk spin axis. 
The hot inner flow inside a truncated disk exhibits 
higher polarization degree at lower energies than 
the sandwich corona.
We interpret this difference as follows:
for the sandwich corona, the first scatterings of 
photons coming from the accretion disk and scattering 
towards the observer create a net polarization parallel to the 
accretion disk that competes with the perpendicular polarization of the emission scattering multiple times in the plane of the 
corona. 
In contrast, the first scatterings of truncated disk photons entering the hot inner flow from the sides create a net perpendicular polarization similar to the perpendicular 
polarization of the photons scattering multiple times 
in the plane of the hot flow.
In principle, high-precision polarization measurements 
can distinguish between the two models. 
However, the uncertainties about the shape and properties of the corona and the disk preclude us from drawing firm 
conclusions.  

The polarization degree of the observed keV photons
are higher if the corona Compton scatters 
synchrotron photons (rather than accretion disk photons).
In this case, $\sim$4\% polarization degrees can already
be observed for $i\ge45^{\circ}$ (Figure \ref{f:c1}).
As the synchrotron photons initially have lower energies
($\sim$1--10~eV) than the accretion disk photons ($\sim$0.1\,keV), more scatterings are required to
scatter them into the keV energy range, leading to
high but rather constant 2-8 keV polarization degrees.

Figure \ref{f:c2} shows the simulation results for models with coronae located on the spin axis of the accretion disk.
The cone shaped corona simulated with \textsc{kerrC} includes the
effects of the reflected emission and exhibits small 
($<2\%$) 2--8 keV polarization degree for $i=30\degr$ and 
$i=45\degr$ inclinations. For $i=60\degr$, the polarization of
the emission from the corona reaching the observer directly, and the emission from the corona reflecting off the disk cancel to give
$\lesssim1\%$ polarization degree at all energies. 
For $i=75\degr$, the polarization parallel to the disk 
is higher, giving a net polarization was calculated reaching $\sim$3\%.
Although even larger inclination can produce polarization degree meeting
or exceeding the observed 4\% polarization degree, the direction stays
parallel to the disk, contradicting the observed alignment of the 
polarization direction and the radio jet.
The polarization of the \textsc{Monk} extended lamppost model (including the effect of the reflected emission)
was calculated  for $a=0$ and $a=0.998$, respectively.
The high-spin models exhibit polarization degree meeting
or exceeding the observed 4\% polarization degree but again, the polarization direction is parallel to the accretion disk.

\subsubsection*{Optical polarimetry}

The optical polarimetric observations were performed using DIPol-2 polarimeter, installed on the remotely operated Tohoku 60 cm (T60) telescope at the Haleakala Observatory, Hawaii. 
DIPol-2 is a double-image CCD polarimeter, capable of measuring linear and circular polarization in three ($B$, $V$, and $R$) 
optical filters simultaneously \cite{PBB14,Piirola21}. 
The design of this instrument optically eliminates the sky polarization (even if it is variable) to a 
polarization level of $<10^{-5}$. 
The instrumental polarization is $<10^{-4}$ and measured by observing twenty unpolarized nearby stars. 
The zero point of the polarization angle 
was determined by observing two highly polarized standard stars (HD\,20 4827 and HD\,25 443).
We observed Cyg X-1 for five nights during the week 2022 May 15 to 21, 
for about 4 hours each night. 
Each measurement of Stokes parameters took about 20~s and we obtained 2298  simultaneous measurements of the normalized Stokes parameters
$q_{\rm obs}=Q_{\rm obs}/I_{\rm obs}$ and 
$u_{\rm obs}=U_{\rm obs}/I_{\rm obs}$ in the three filters ($B$, $V$, and $R$).
These individual measurements were used to compute average intranight values of Stokes parameters using the $2\sigma$ weighting algorithm \cite{Kosenkov17,Piirola21}.
The uncertainty of the final average corresponds to the standard deviation of individual measurements resulting from the orbital variability of the source.
The polarization produced by the interstellar (IS) medium was estimated by observing a sample of field stars (Figure~\ref{fig:field_map}), which are close in distance to the target as indicated by their Gaia parallaxes 
(Figure~\ref{fig:pol_parallax}) \cite{2021A&A...649A...1G,2021A&A...649A...4L}.
Taking into account angular separation on the image, closeness in distance, and the wavelength dependence of the polarization, we choose two stars (designating them Ref 1 and Ref 2) from our sample as the IS polarization standards (see Figure~\ref{fig:field_map}). 
We considered two cases: the Stokes parameters of the IS polarization were set to be equal to those of Ref 2,
and, alternatively, to the weighted average of those of Ref 1 and Ref 2. 
For both cases, the normalized Stokes parameters ($q_{\rm is}$, $u_{\rm is}$) were subtracted from the measured values of Stokes parameters of the target ($q_{\rm obs}$, $u_{\rm obs}$) to obtain the intrinsic polarization ($q_{\rm int}$, $u_{\rm int}$) estimates.
From this we determine the intrinsic polarization degree (PD) 
and polarization angle (PA)  as
\begin{equation} \label{eq:PDPA}
 \mbox{PD} =  \sqrt{q_{\rm int}^2+u_{\rm int}^2}, \quad 
 \mbox{PA}= \frac{1}{2} {\rm atan}2 (u_{\rm int},q_{\rm int}).
\end{equation} 
The uncertainty on the polarization degree $\Delta \mbox{(PD)}$ was estimated as the uncertainty of the individual Stokes parameters, and includes both the source and IS polarization uncertainties.
The uncertainty on the polarization angle (in radians) was estimated as $\Delta \mbox{(PA)} =  \Delta \mbox{(PD)} /(2\,\mbox{PD})$ \cite{Serkowski62}.
The observed normalized Stokes parameters, the IS polarization and the intrinsic Stokes parameters as well as the polarization degree and polarization angle are reported in Table~\ref{tab:dipol}. 

We used the RoboPol polarimeter in the focal plane of the 1.3\,m telescope of the Skinakas observatory (Greece) to 
obtain additional $R$-band  polarimetry.
The observations were performed between 2022 May 13 and June 2 with multiple pointings in 10 nights.
In total, 21 exposures series were acquired, 
each series consisting of 10 to 20 exposures, each of 1 to 2 seconds duration.
The instrumental polarization was found with a set of unpolarized standards stars (BD\,+28\,4211, BD\,+33\,2642, BD\,+32\,3739, BD\,+40\,2704, HD\,154\,892). 
The zero polarization angle was determined based on three highly polarized standard stars (VI Cyg 12, Hiltner 960 and CygOB2 14). 
The Cyg X-1 measurements do not reveal any polarization variability exceeding that of the standard stars (for which the standard deviation from the mean values, $\sigma_q=0.12\%, \sigma_u=0.08\%$, were obtained). 
We determined the average polarization parameters of Cyg X-1 from calculating the sigma-clipped median of the relative Stokes parameters. 
The uncertainties were determined by error propagation adding the instrumental polarization uncertainties in quadrature.  We determined the intrinsic source 
polarization by subtracting the IS polarization 
using the same Ref 2 star as used in the DIPol-2 analysis (Table~\ref{tab:dipol}).

We find optical polarization angles of Cyg X-1 between
$-37\degr$ to $-11\degr$, close to the position angle 
of the jet from radio interferometry
(from $-26\degr$ to $-9\degr$)  \cite{2001MNRAS.327.1273S,2011ApJ...742...83R}. The blue supergiant companion star
dominates the optical emission from Cyg X-1 \cite{2014MNRAS.442.3243Z}.
The optical polarization is likely produced by the
scattering of the stellar radiation off the bulge 
formed by the accretion stream interacting with 
the accretion disk \cite{Kemp1979}.
\clearpage 
\renewcommand{\thefigure}{{\bf S1}}
\begin{figure}
\begin{center}
\vspace*{-1cm}
    \includegraphics[width=14cm]{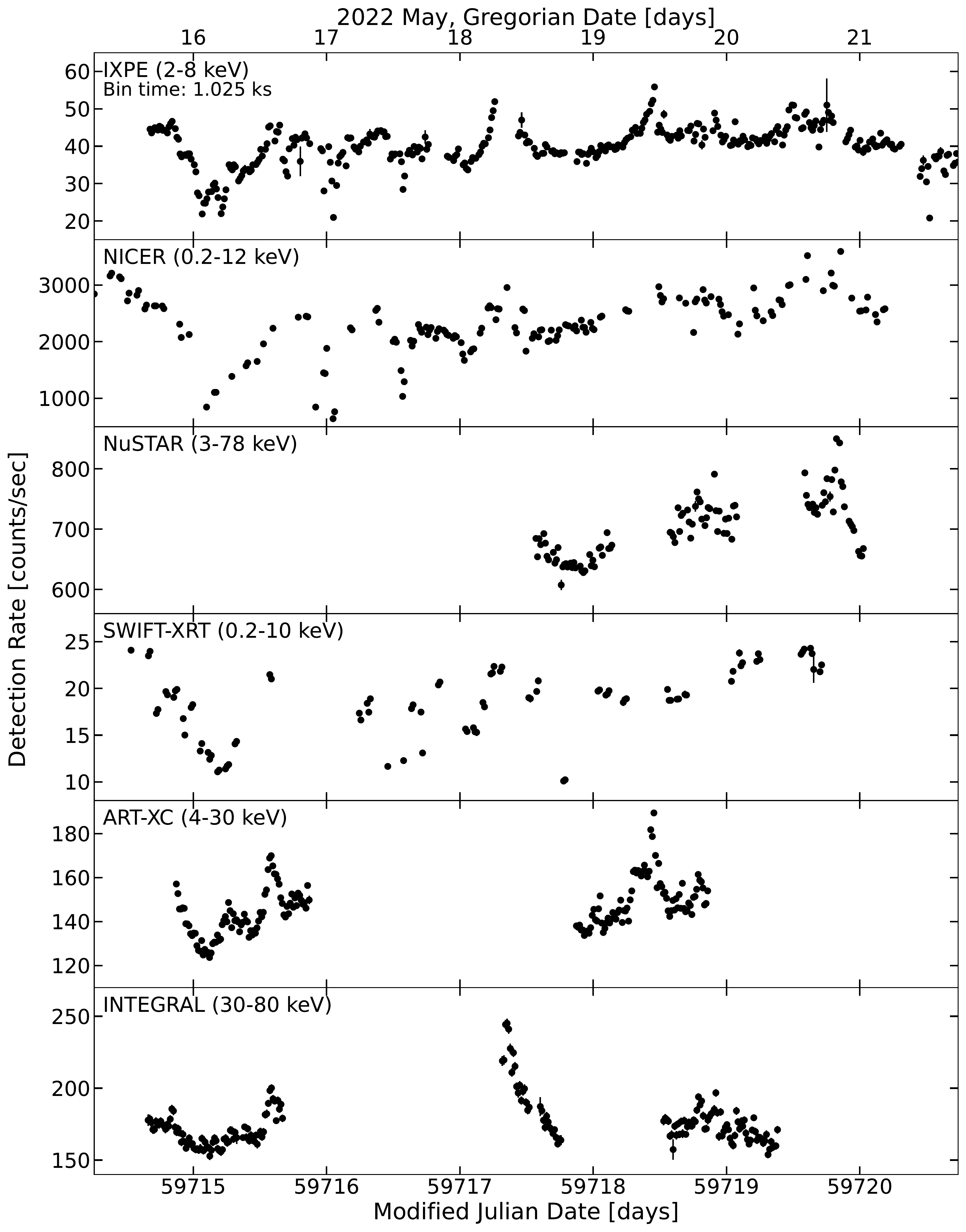}
\vspace*{-0.5cm}
    \end{center}
    \caption{\textbf{X-ray light curves of Cyg X-1 from the 2022 May 15 to 21 observation campaign.} 
    From top to bottom: the IXPE, NICER, NuSTAR,
    Swift/XRT, SRG/ART-XC, and INTEGRAL light curves.
\label{f:lightcurves}}
\end{figure}

\renewcommand{\thefigure}{{\bf S2}}
\begin{figure}
\begin{center}
    \includegraphics[width=\textwidth]{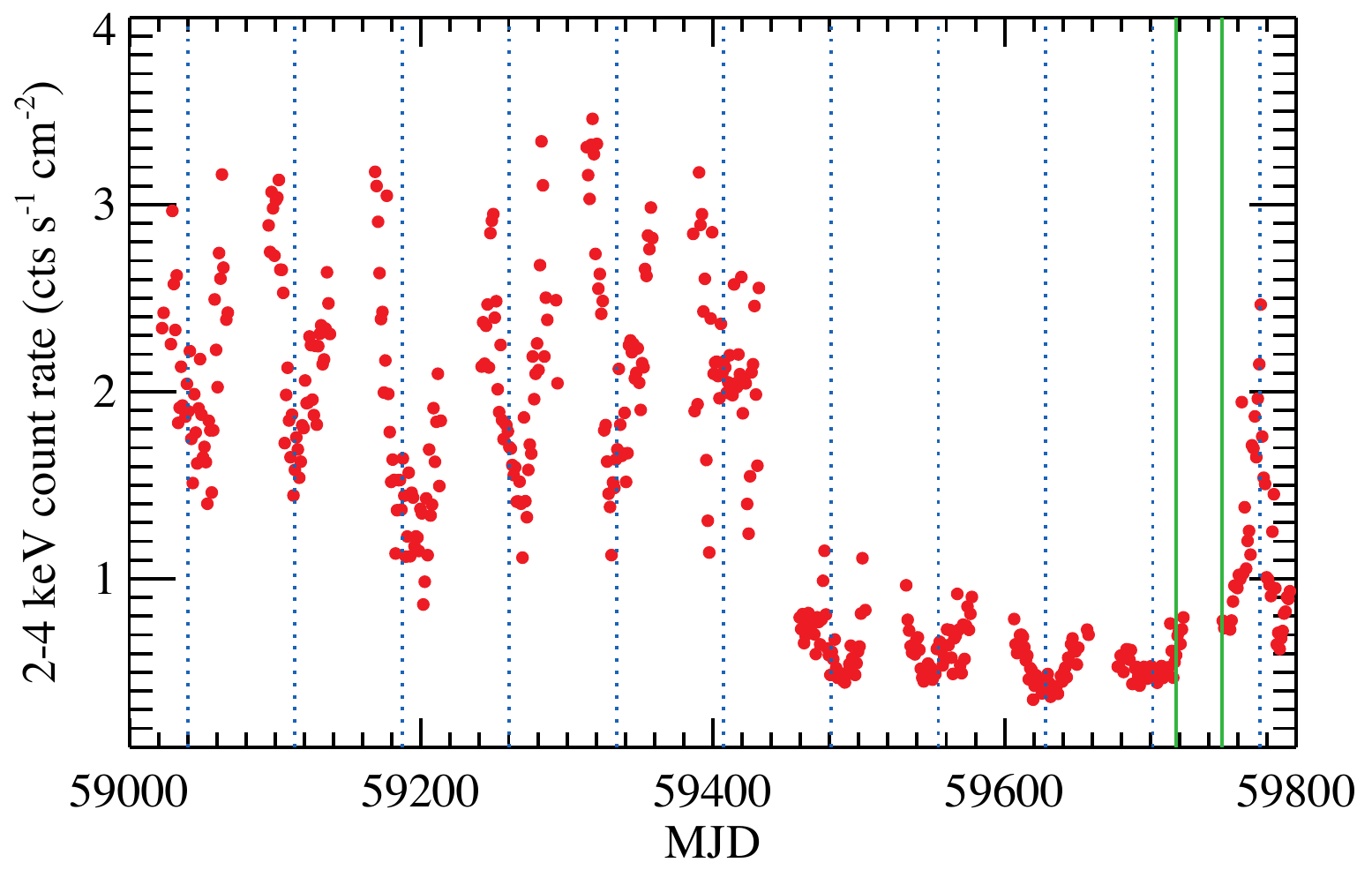}
\end{center}
    \caption{\textbf{Long-term Cyg X-1 x-ray light curve.}
    The figure shows the daily 2--4 keV count rate obtained from the MAXI monitor from May 31, 2020 (MJD 59000) 
    to August 9, 2022 (MJD 59800).
    Phases of high 2--4 keV fluxes during the soft state and low 2--4 keV fluxes during the hard state can be recognized.
    The vertical dotted lines (blue) show the 
    dates of the superorbital flux minima, appearing at
    ${\rm MJD} = 59040.0 + 73.5 n$, 
    with $n$ being an integer number.
    The two vertical solid lines (green)
    show the mid-times of two IXPE campaigns, 2022 May 15 to 21 and June 18 to 20, respectively.
    The first observation was close to the 
    superorbital flux minimum, and the second 
    was shifted by about half-period.
    The second observation was taken right before the short incursion into the soft state.
\label{f:sop}}
\end{figure}

\renewcommand{\thefigure}{{\bf S3}}
\begin{figure}
\begin{center}
    \includegraphics[totalheight=12cm]{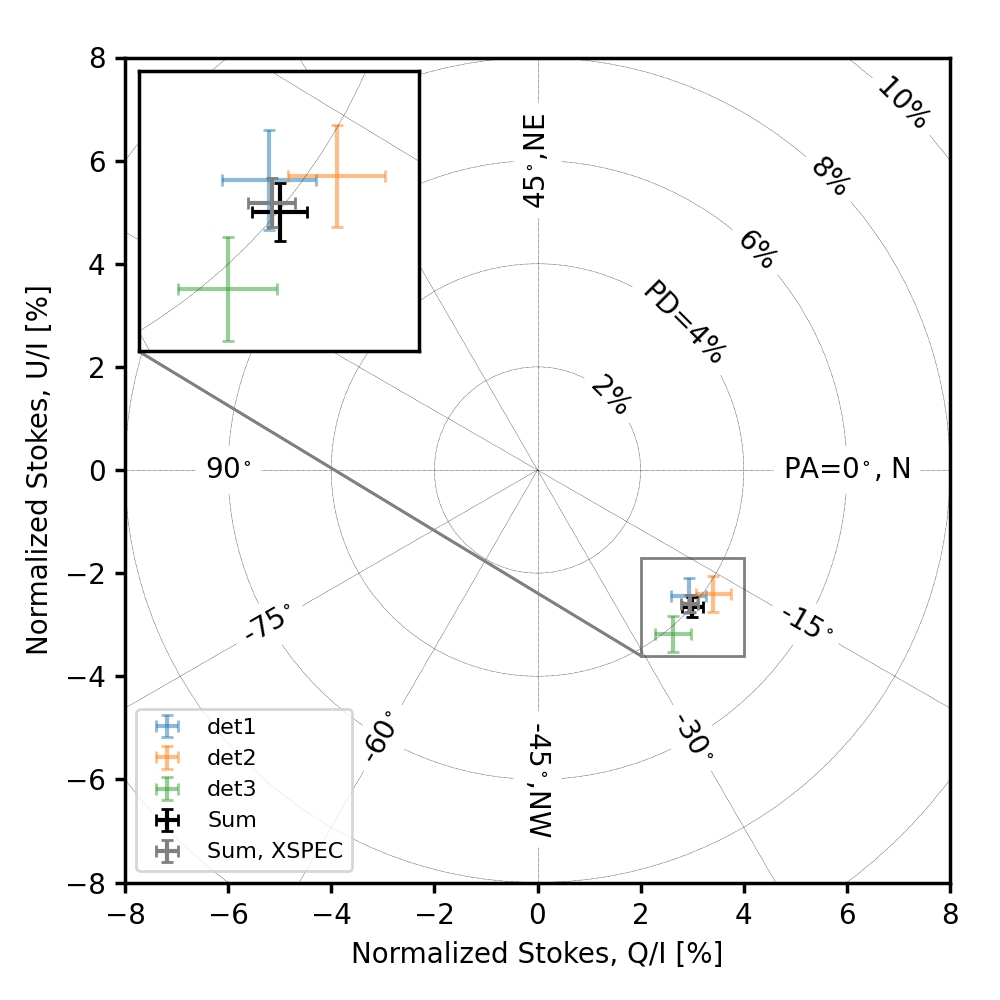}\hspace*{1cm}
\end{center}
    \caption{\textbf{X-ray linear polarization of Cyg X-1 from the 2022 May 15 to 21 observations.}
    The linear polarization of the x-rays from Cyg X-1 is shown in the plane of the normalized 
    Stokes $Q/I$ and $U/I$ parameters measured with each of the three IXPE x-ray telescopes (coloured data points), and for the combined signal from all three telescopes (black). The grey data point shows the
    results from the analysis of the data using the \xspec tool, 
    instead of \ixpeobssim. The two approaches give a result which is compatible within the statistical uncertainties.
    The circles give the contours of constant polarization degree (PD) while the radial lines correspond to constant polarization angle (PA). The error bars are 1\,$\sigma$.
     \label{f:stokes}}
\end{figure}

\renewcommand{\thefigure}{{\bf S4}}
\begin{figure}
\begin{center}
    \includegraphics[totalheight=12cm]{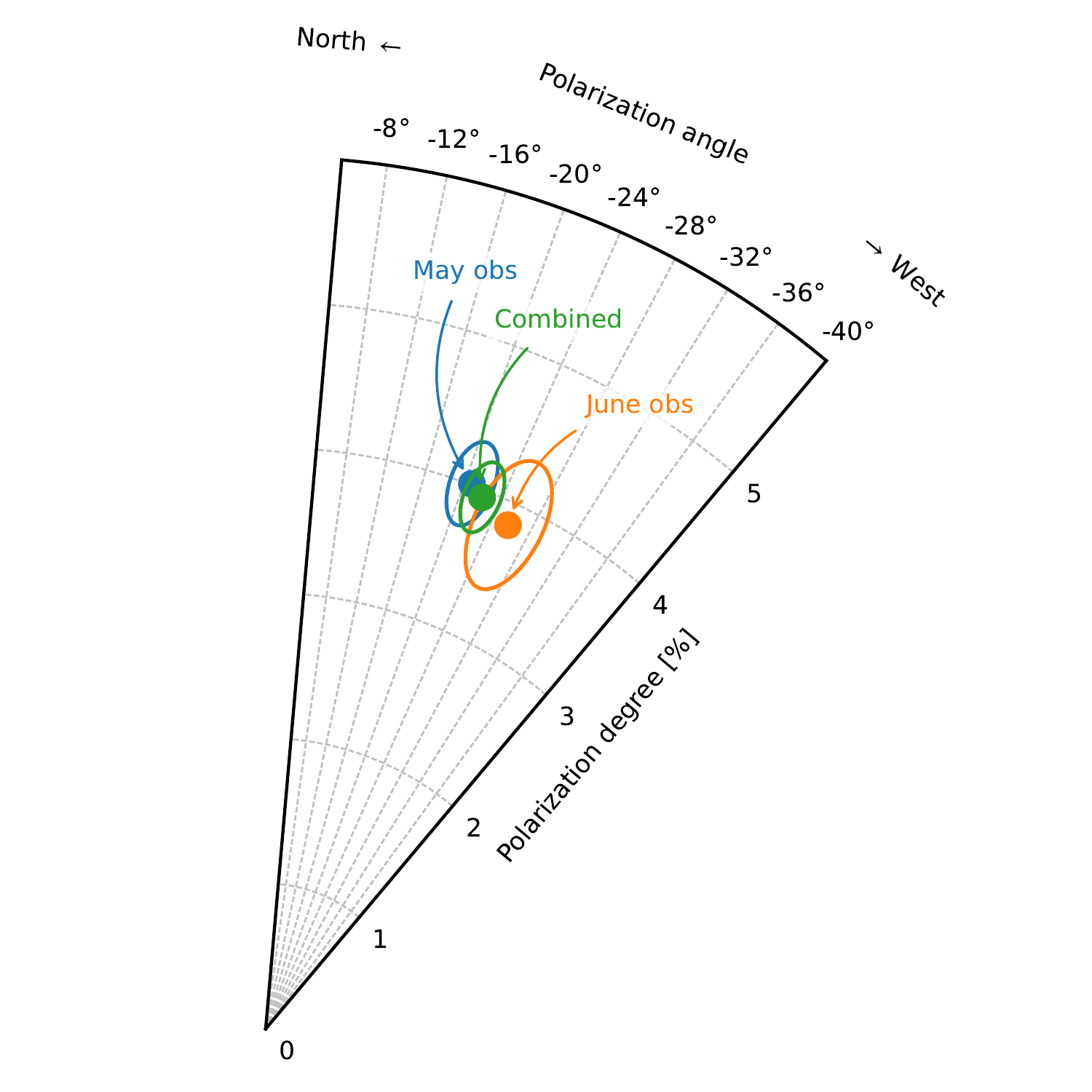}\hspace*{1cm}
\end{center}
    \caption{\textbf{Linear x-ray polarization of Cyg X-1 measured in two occasions, as well as the combined result.} The figure shows the polarization degree and angle of the 2022 May 15 to 21 observations (blue), the 2022 June 18 to 20 observations
    (orange), and for the combined data set (green).
    For each result the most likely values (circles) and 68.3\% confidence regions (ellipses) 
    are shown. 
     \label{f:mj}}
\end{figure}

\renewcommand{\thefigure}{{\bf S5}}
\begin{figure}
\begin{center}
    \includegraphics[width=0.6\columnwidth]{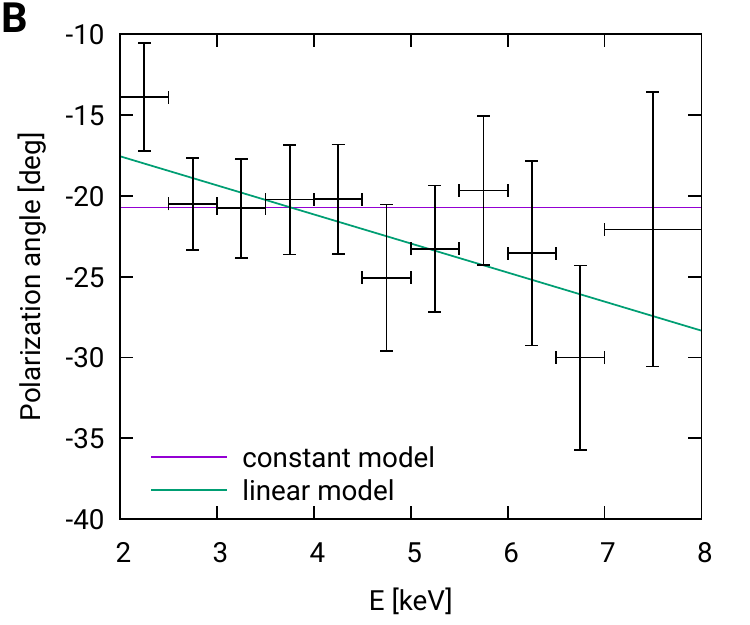}\\[3mm]
    \includegraphics[width=0.6\columnwidth]{Figure_S5a.pdf}
\end{center}
    \caption{\textbf{Energy dependence of the observed polarization degree (A) and polarization angle (B).} The data (black crosses with 1$\sigma$ error bars) are produced using the PCUBE algorithm of the {\tt xpbin} tool and summed over all detector units. The constant (violet) and linear (green) models fitted to the data are also depicted (see the text for details).
     \label{f:pd-fit}}
\end{figure}

\renewcommand{\thefigure}{{\bf S6}}
\begin{figure}
\centering
\includegraphics[width=0.6\columnwidth]{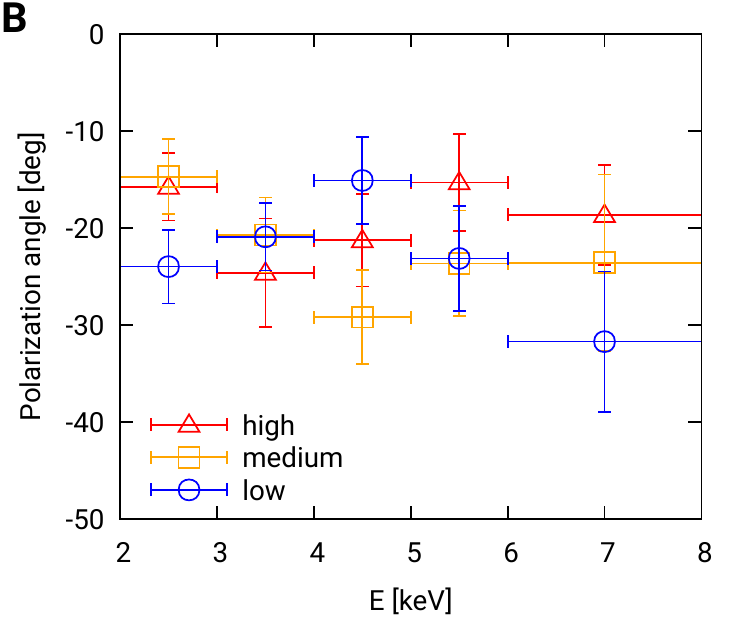}\\[3mm]
\includegraphics[width=0.6\columnwidth]{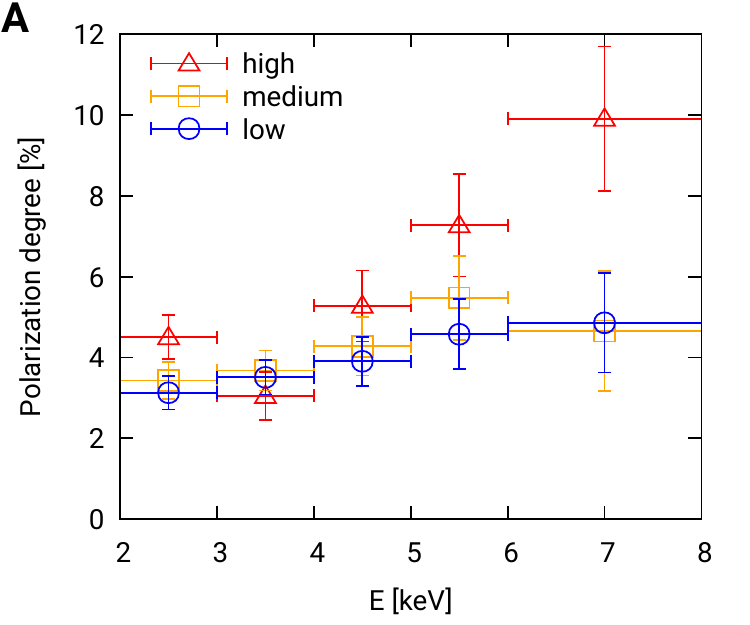}
\caption{\textbf{Polarization of Cyg X-1 at different flux levels.} Comparison of the polarization degree \textbf{(A)} and polarization angle \textbf{(B)} for three different flux selected data sets.  
\label{fig:polfracang_brightdip}}
\end{figure}

\renewcommand{\thefigure}{{\bf S7}}
\begin{figure}
\centering
\includegraphics[width=1.\columnwidth]{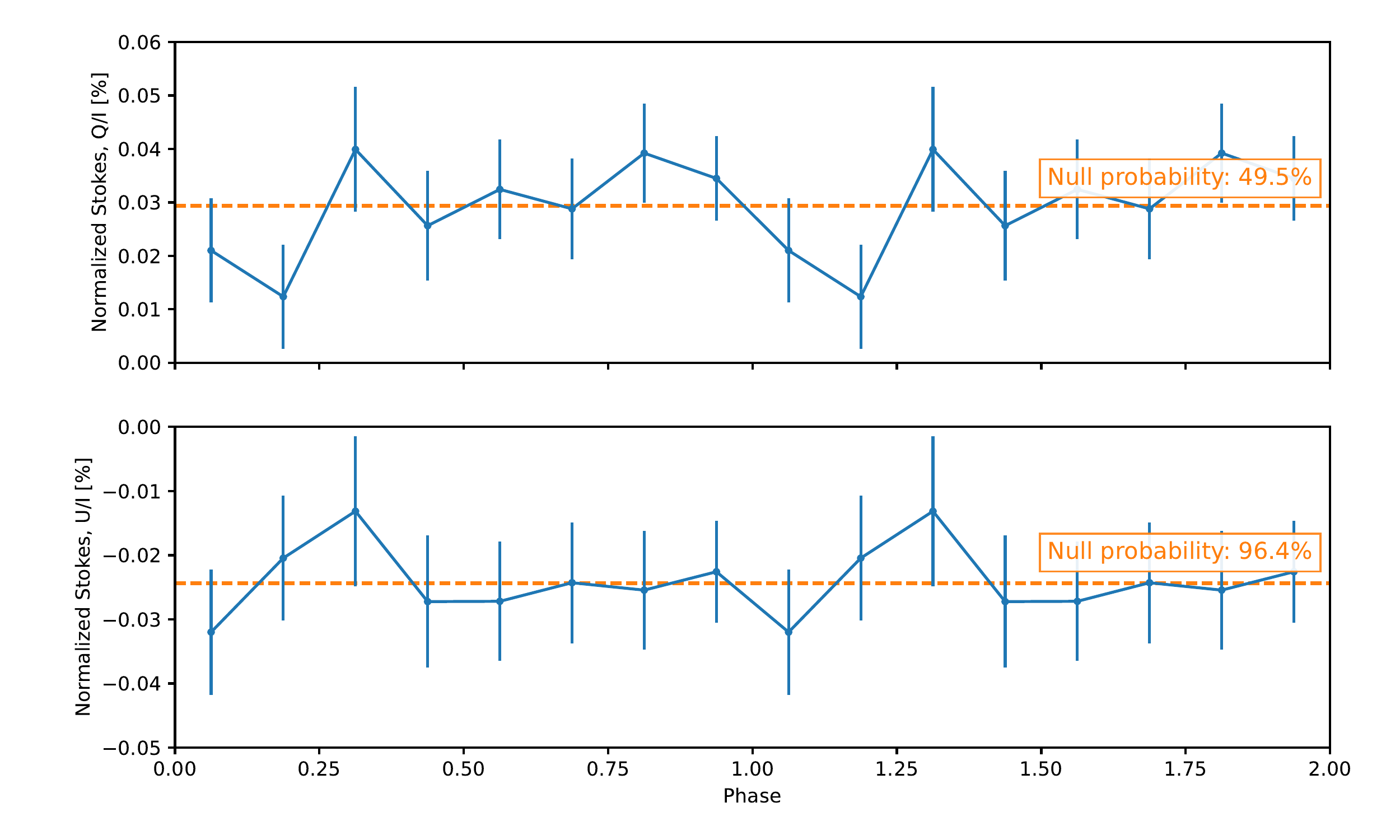}\\[3mm]
\caption{\textbf{Orbital phase dependence of the Cyg X-1 x-ray polarization properties.} 
The observed x-ray normalized Stokes parameters $Q/I$ and $U/I$ (summed from 2 to 8 keV) 
are statistically consistent with being constant
as a function of the orbital phase.
Note that the results are shown for two orbital
periods. 
The orbital phase of 0 corresponds to the 
superior conjunction maximizing the stellar wind
absorption of the x-rays. 
\label{f:op}}
\end{figure}

\linespread{1.0}

\renewcommand{\thefigure}{{\bf S8}}
\begin{figure}
\begin{center}
\vspace*{-2cm}
\includegraphics[totalheight=15.5cm]{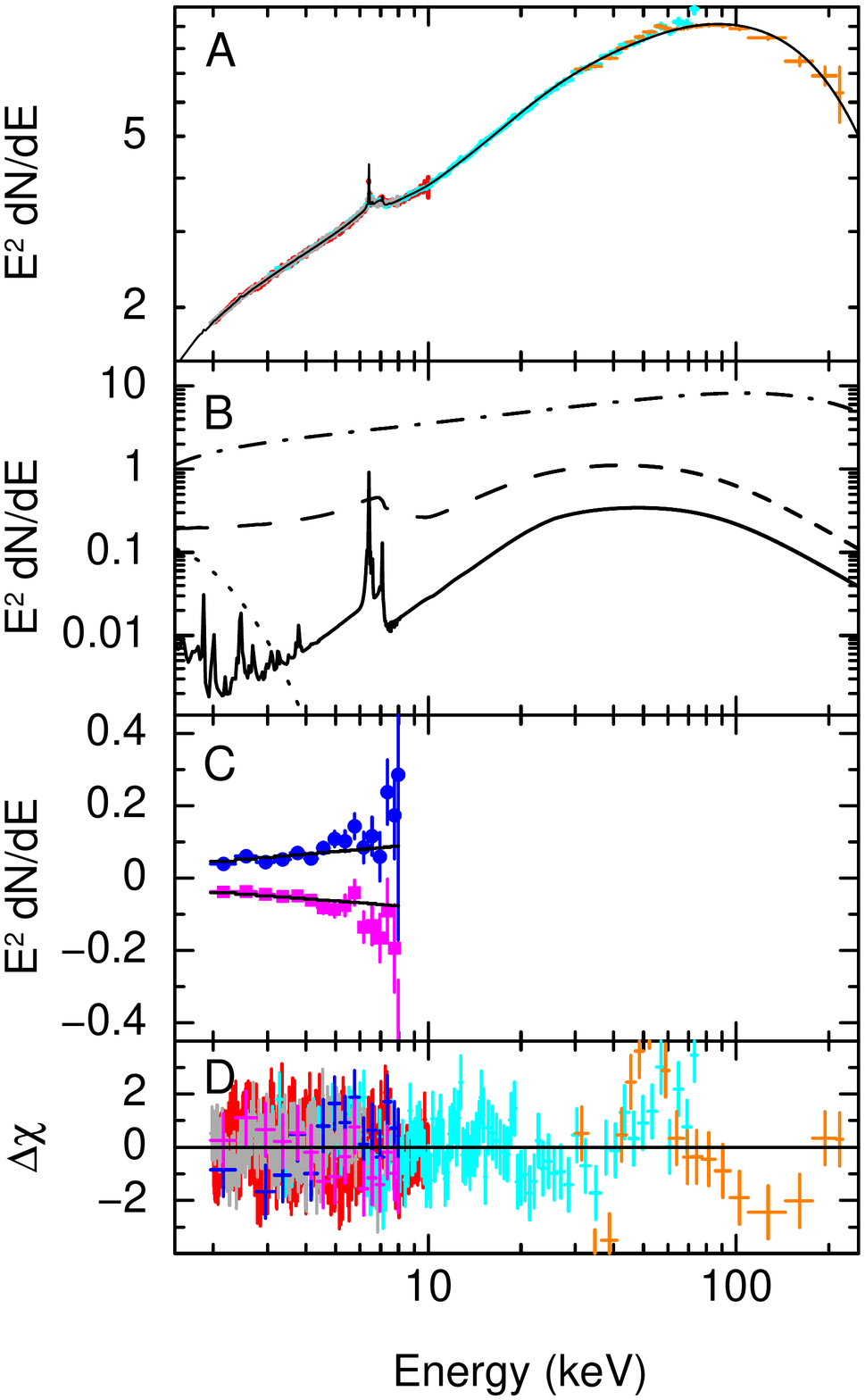}
    \vspace*{-1cm}
    \end{center}
    \caption{\textbf{Results of spectropolarimetric fitting.} \textbf{(A)}  NICER (red), NuSTAR (cyan), IXPE (grey) and INTEGRAL/ISGRI (orange) Stokes $I$ spectrum unfolded around the best-fit model (black solid line). 
    For each bin of the energy spectrum, the unfolded data point is
    the number of observed counts times the best-fit model value 
    divided by the counts expected in the bin for the best-fit model.
    For plotting purposes only, data and model are both divided by the relevant \textsc{mbpo} model to remove calibration discrepancies. The specific photon flux $dN/dE$ has units of photons cm$^{-2}$ s$^{-1}$ keV$^{-1}$. \textbf{(B)} Individual components of the best-fit model: thermal disk emission (dotted line),
    Compton scattered emission from the corona (dashed dotted line), relativistic reflection (dashed line), non-relativistic reflection (solid line).
    \textbf{(C)} Stokes $Q$ (blue circles) and $U$ (magenta squares), also unfolded around the best-fit model. \textbf{(D)} Residuals (contributions to $\chi$). For plotting purposes only, data from different detectors of the same observatory have been grouped together, and a maximum of 10 energy channels have been grouped together to achieve a signal-to-noise ratio of 150.}
     \label{f:eeuf01}
\end{figure}

\renewcommand{\thefigure}{{\bf S9}}
\begin{figure}
\begin{center}
\hspace*{-0.7cm}
\includegraphics[width=9.2cm]{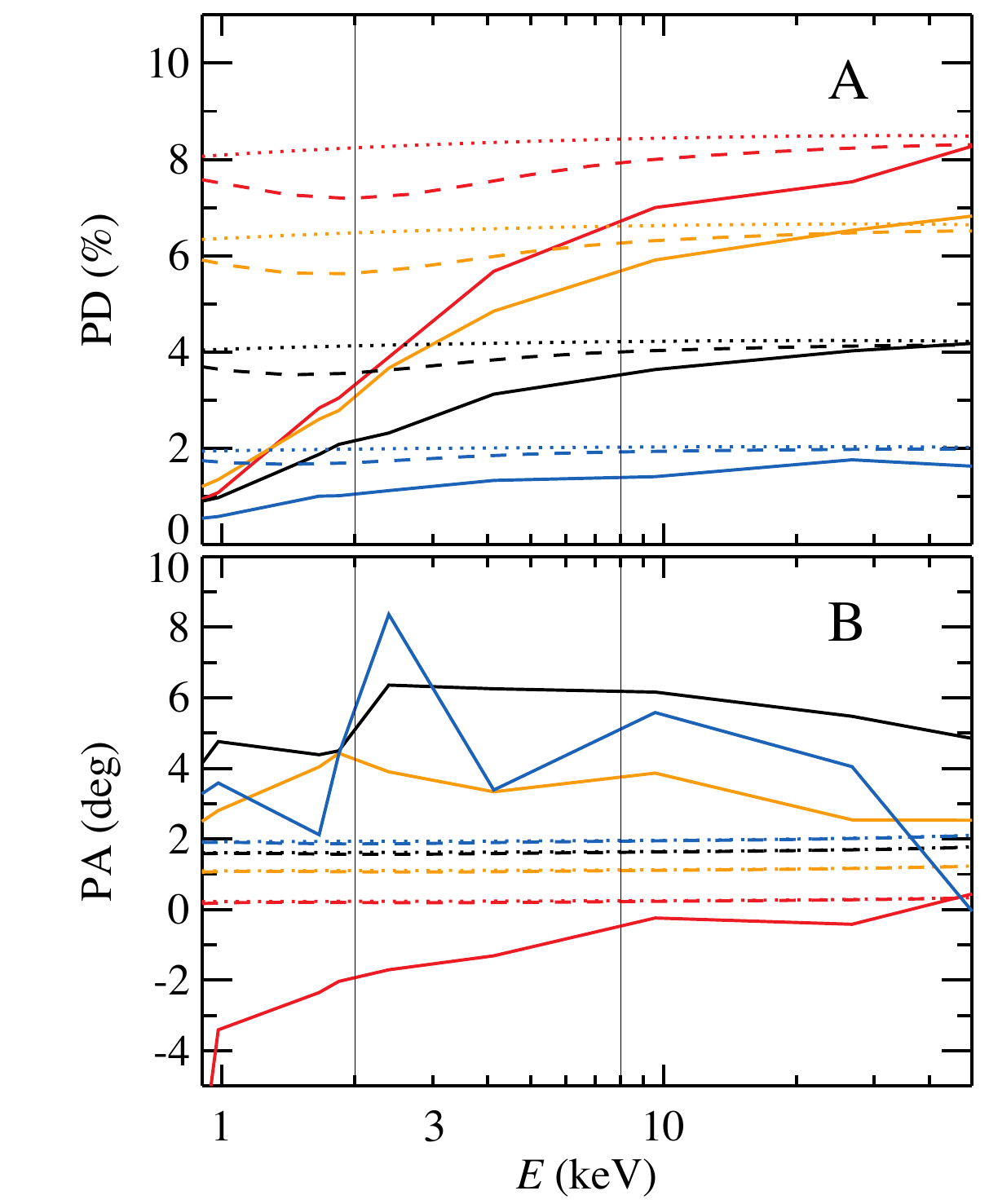}
\vspace*{-0.5cm}
    \end{center}
    \caption{
    \textbf{Polarization degree (A) and polarization angle (B)
    for models with coronae extending parallel to the accretion disk.}  
    The solid lines show the predictions of the
    sandwich corona, the dashed and dotted lines show the predictions of the hot inner flow inside a truncated disk, with accretion disk photons (dashed lines) and synchrotron photons (dotted lines) acting as seed photons for the inverse Compton scattering. 
    The colors encode the inclination angle at which the coronae are observed: red (75\degr), orange (60\degr), black (45\degr) and blue (30\degr). 
    The vertical lines delineate the IXPE band from 2--8 keV.
    For very low polarization degrees the polarization angle in the 
    sandwich corona model fluctuates by a few degrees owing to the finite
    number of simulated events.
    Positive polarization angles correspond to counterclockwise rotations of the polarization vector relative to the projected disk spin axis on the plane of the sky in Figure~\ref{f:comp}.
    \label{f:c1}}
\end{figure}

\renewcommand{\thefigure}{{\bf S10}}
\begin{figure}
\begin{center}
\hspace*{-0.7cm}
\includegraphics[width=9.2cm]{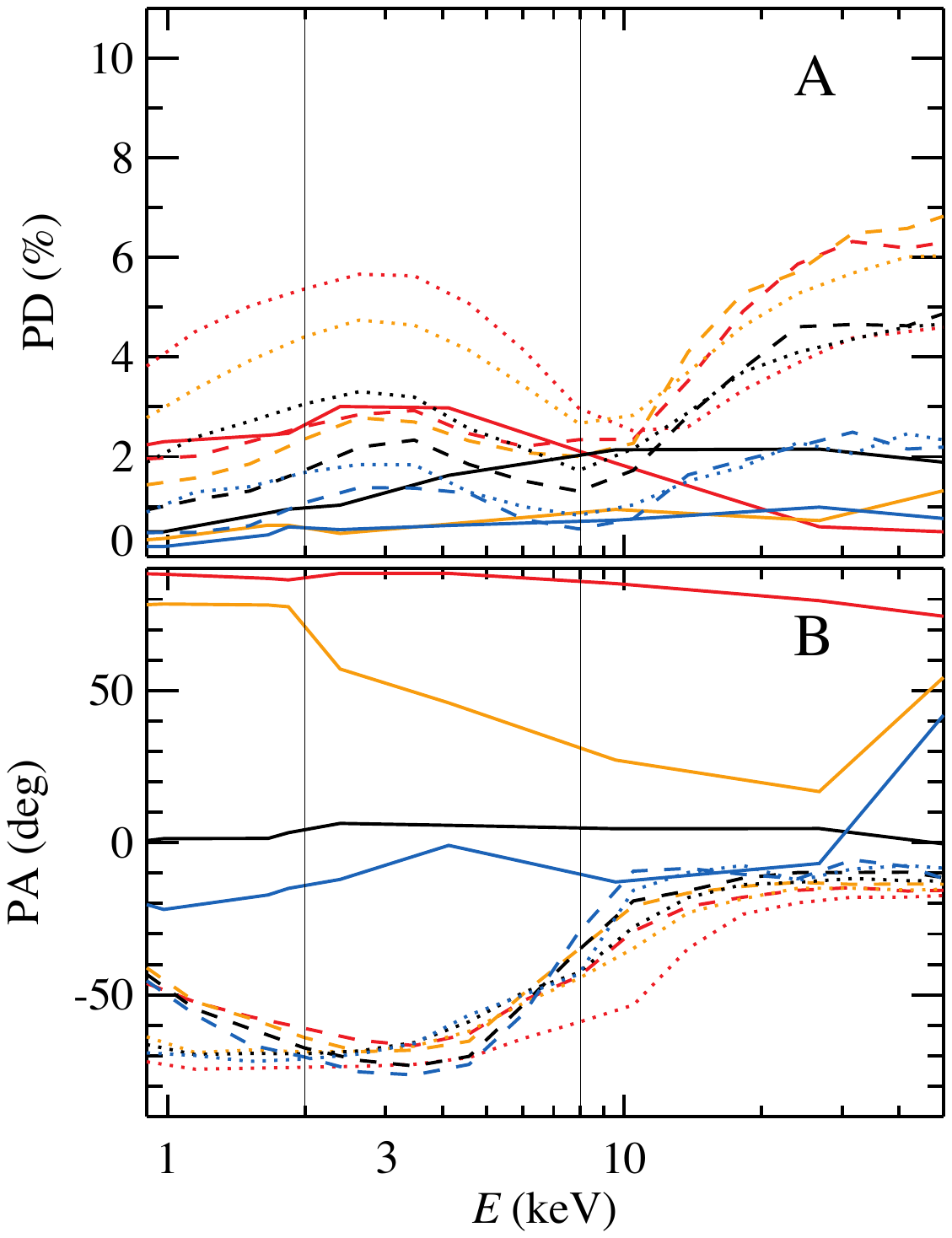}
\vspace*{-0.5cm}
    \end{center}
    \caption{
    \textbf{Same as Figure~\ref{f:c1}, but
    for models with coronae located on the spin axis of the accretion disk.}
    The solid lines show the predictions for a cone-shaped corona
    extended along the disk spin axis, the dashed and dotted lines 
    shows the results for an extended lamppost corona for a 
    non-spinning black hole ($a=0$, dashed line) and a spinning black hole
    ($a=0.998$, dotted line).
    \label{f:c2}}
\end{figure}

\renewcommand{\thefigure}{{\bf S11}}
\begin{figure}
\centering
\includegraphics[width=0.5\columnwidth]{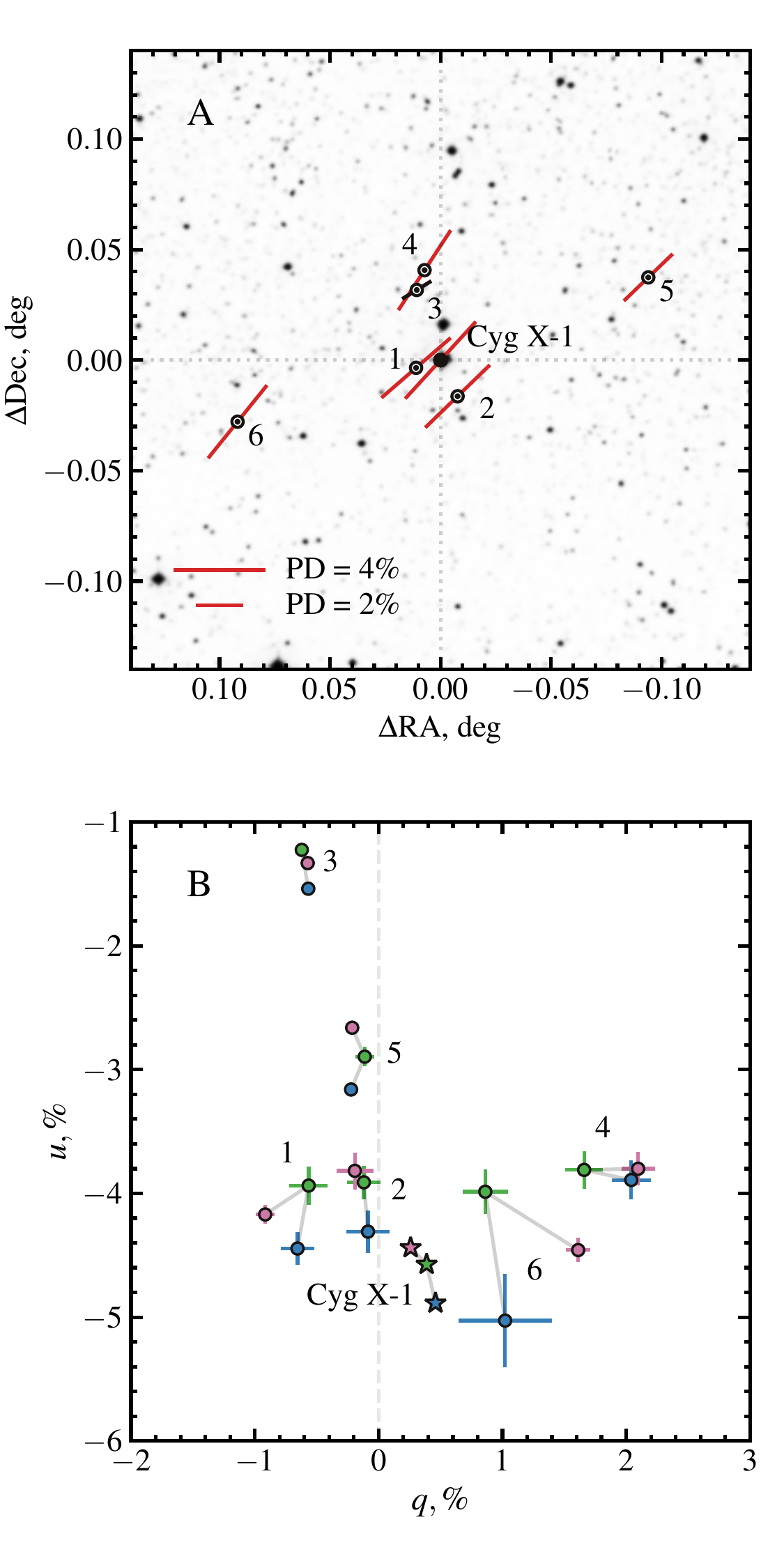}
\caption{\textbf{Polarization of nearby field stars around Cyg X-1.}  
  \textbf{(A)} Polarization vectors of the field stars (open circles) and Cyg X-1 (filled circle) in the $B$-filter, with field stars image as a background. The length of the solid lines is proportional to the polarization degree. 
  The deviations in declination ($\Delta$Dec) and right ascension 
  ($\Delta$RA) are relative to the Cyg X-1 position (grey 
  dotted lines).
  \textbf{(B)} The observed normalized Stokes parameters $q$ and $u$ for the field stars (circles) and Cyg X-1 (stars). Blue, green and magenta colors correspond to $B$, $V$, and $R$ filters, respectively. 
  For clarity, the grey solid lines connect the $B$, $V$, and $R$ 
  results for each source. Uncertainties are 1$\sigma$. The vertical grey dashed line indicates the $q=0$ axis.
  Stars Ref 1 and Ref 2 are chosen as the IS polarization standards.
\label{fig:field_map}}
\end{figure}

\renewcommand{\thefigure}{{\bf S12}}
\begin{figure}
\centering
\includegraphics[width=0.6\columnwidth]{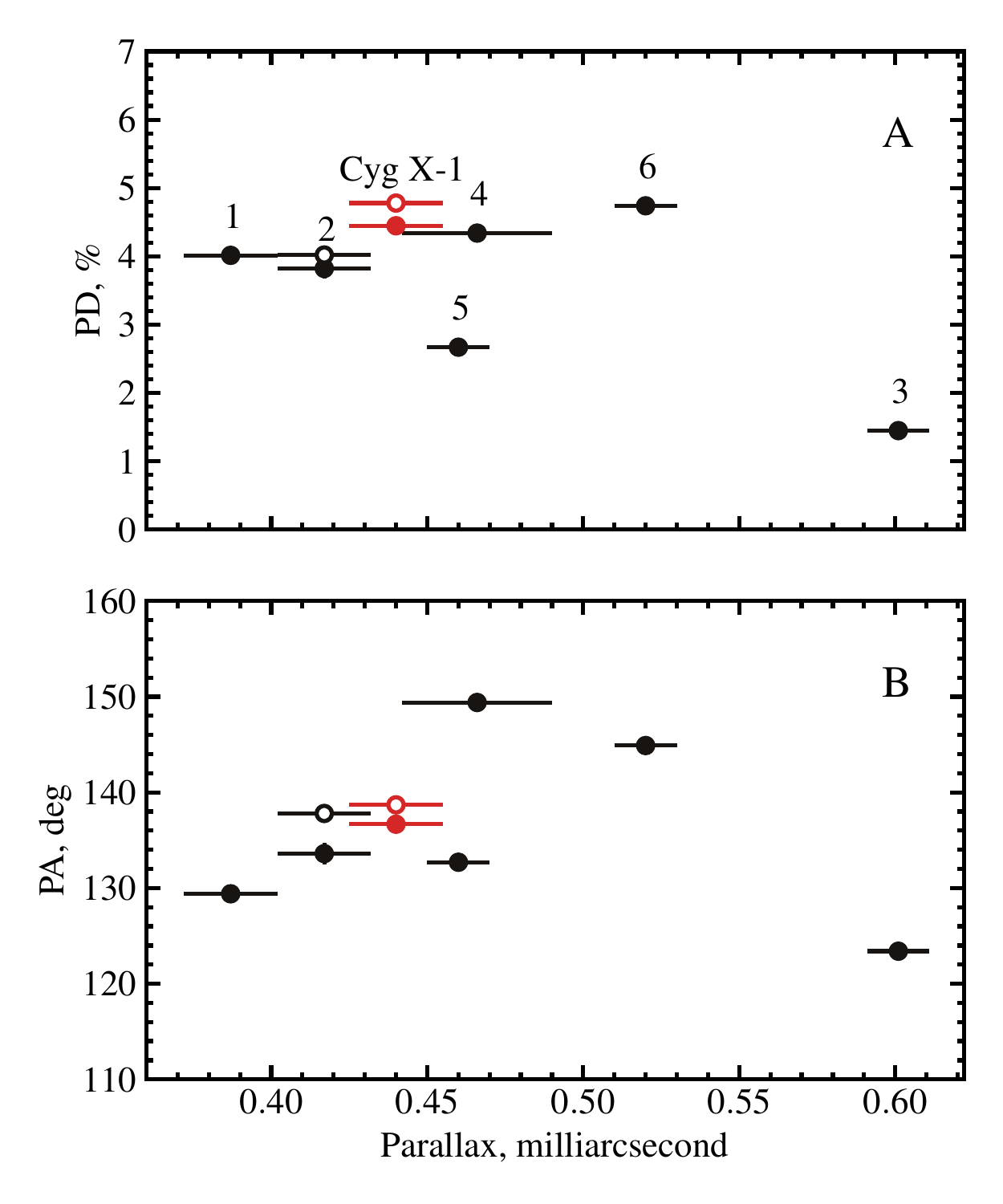}
\caption{\textbf{Polarization of nearby field stars around Cyg X-1 as a function of parallax.} 
\textbf{(A)} Polarization degree (PD) and  \textbf{(B)} polarization angle (PA) for a set of field stars (black) and Cyg X-1 (red) as measured with DIPol-2 (filled circles) and RoboPol (open circles) in the $R$-band. 
Error bars show uncertainties at the 1$\sigma$ confidence level. 
\label{fig:pol_parallax}}
\end{figure}

\clearpage
\renewcommand{\thetable}{{\bf S1}}
\begin{table}
\caption{\textbf{IXPE polarization results given in terms of the Stokes parameters.} Values are derived for the data collected independently by each individual IXPE telescopes and for their sum with the \ixpeobssim and (only for the sum) with \xspec analysis. The two methods and the independent analysis of single IXPE telescopes provide consistent results. The uncertainties are 68.3\% confidence interval, assuming that the Stokes parameters are independent.}
\begin{tabular}{lrrrrr}
\hline
\hline
{} &           2.0--3.0 keV &           3.0--4.0 keV &           4.0--6.0 keV &           6.0--8.0 keV &           2.0--8.0 keV \\
\hline
$Q/I$ - det1 [\%]         & $2.9\pm 0.5$ & $2.5\pm 0.5$ & $3.4\pm 0.6$ &  $3.0\pm1.4$ &   $2.9\pm 0.3$ \\
$Q/I$ - det2 [\%]         & $3.3\pm 0.5$ & $2.5\pm 0.5$ & $3.9\pm 0.6$ &  $4.7\pm1.4$ &   $3.4\pm 0.3$ \\
$Q/I$ - det3 [\%]         & $2.1\pm 0.5$ & $2.7\pm 0.5$ & $3.1\pm 0.6$ &  $3.4\pm1.6$ &   $2.6\pm 0.4$ \\
$Q/I$ - sum [\%]          & $2.8\pm 0.3$ & $2.5\pm 0.3$ & $3.5\pm 0.3$ &  $3.7\pm0.8$ &   $3.0\pm 0.2$ \\
$Q/I$ - sum (\xspec) [\%] & $2.9\pm 0.3$ & $2.7\pm 0.3$ & $3.4\pm 0.3$ &  $3.7\pm0.8$ &   $2.9\pm 0.2$ \\
\hline
$U/I$ - det1 [\%]         & $-2.1\pm 0.5$ & $-1.9\pm 0.5$ & $-2.8\pm 0.6$ & $-4.2\pm1.4$ &  $-2.4\pm 0.3$ \\
$U/I$ - det2 [\%]         & $-1.3\pm 0.5$ & $-2.3\pm 0.5$ & $-2.7\pm 0.6$ & $-6.0\pm1.4$ &  $-2.4\pm 0.3$ \\
$U/I$ - det3 [\%]         & $-2.9\pm 0.5$ & $-2.9\pm 0.5$ & $-4.0\pm 0.6$ & $-2.9\pm1.6$ &  $-3.2\pm 0.4$ \\
$U/I$ - sum [\%]          & $-2.1\pm 0.3$ & $-2.3\pm 0.3$ & $-3.1\pm 0.3$ & $-4.5\pm0.8$ &  $-2.7\pm 0.2$ \\
$U/I$ - sum (\xspec) [\%] & $-2.3\pm 0.3$ & $-2.4\pm 0.3$ & $-3.2\pm 0.3$ & $-4.2\pm0.8$ &  $-2.6\pm 0.3$ \\
\hline
\end{tabular}
\label{tab:Stokes}
\end{table}

\renewcommand{\thetable}{{\bf S2}}
\begin{table}
\caption{\textbf{IXPE polarization results given in terms of the polarization degree and angle.} Uncertainties are given on 68.3\% confidence level, 
and were calculated from the Stokes parameters reported 
in Table~\ref{tab:Stokes} assuming that the polarization 
degree and polarization angle are independent.
The significance was calculated as the measured polarization degree divided by the uncertainty, for the sum of the three IXPE telescopes. 
}
\begin{tabular}{llllll}
\hline
\hline
{} &          2.0--3.0 keV &          3.0--4.0 keV &          4.0--6.0 keV &          6.0--8.0 keV &          2.0--8.0 keV \\
\hline
PD - det1 [\%]          &  $3.5\pm0.5$ &  $3.1\pm0.5$ &  $4.4\pm0.6$ &  $5.1\pm1.4$ &  $3.8\pm0.3$ \\
PD - det2 [\%]          &  $3.6\pm0.5$ &  $3.4\pm0.5$ &  $4.7\pm0.6$ &  $7.6\pm1.4$ &  $4.2\pm0.3$ \\
PD - det3 [\%]          &  $3.6\pm0.5$ &  $3.9\pm0.5$ &  $5.1\pm0.6$ &    $4.5\pm1.6$ &  $4.1\pm0.4$ \\
PD - sum [\%]           &  $3.5\pm0.3$ &  $3.5\pm0.3$ &  $4.7\pm0.3$ &  $5.8\pm0.8$ &  $4.0\pm0.2$ \\
PD - sum (\xspec) [\%]  &  $3.7\pm0.3$ &  $3.6\pm0.3$ &  $4.7\pm0.3$ &  $5.6\pm0.8$ &  $3.9\pm0.2$ \\
PD significance          &          13$\sigma$ &          12$\sigma$ &          14$\sigma$ &           7$\sigma$ &          20$\sigma$ \\
\hline
PA - det1 [deg]         &  $-18\pm4$ & $-19\pm4$ & $-20\pm4$ & $-27\pm8$ &          $-20\pm3$ \\
PA - det2 [deg]         &  $-11\pm4$ & $-22\pm4$ & $-17\pm4$ & $-26\pm5$ &          $-18\pm2$ \\
PA - det3 [deg]         &  $-27\pm4$ & $-23\pm4$ & $-26\pm4$ & $-20\pm10$ & $-25\pm2$ \\
PA - sum [deg]          &  $-18\pm2$ & $-21\pm2$ & $-21\pm2$ & $-25\pm4$ &          $-21\pm1$ \\
PA - sum (\xspec) [deg] &  $-19\pm2$ & $-21\pm2$ & $-21\pm2$ & $-25\pm4$ &          $-21\pm1$ \\
\hline
\end{tabular}
\label{tab:pol}
\end{table}

\renewcommand{\thetable}{{\bf S3}}
\begin{table}
\caption{{\bf Parameters of the \textsc{kerrC} models shown in 
Figures \ref{f:c1} and \ref{f:c2}.}}
\begin{center}
\begin{tabular}{p{8cm}llp{2cm}p{2cm}}
\hline
\hline
Parameter & Symbol & Unit & wedge & cone\\  \hline
Black hole spin & $a$ & none & 0.9 & 0.9\\
Black hole mass & $M$ & solar masses & 
21.2 & 21.2 \\
Corona temperature & $T_{\rm C}$ & keV & 100 & 150\\
Optical depth & $\tau_{\rm C}$ & none & 0.35 &0.79\\
Opening angle & $\theta_{\rm C}$ & deg & 10 &25\\
Corona inner/outer edge & $r_1,r_2$ & $r_{\rm g}$ & 2.32/100 & 2.5/20 \\
Inclination & $i$ & deg & 65 & 85 \\  
Accretion rate & $\dot{M}$ & 10$^{18}$ g~s$^{-1}$ & 0.0505 & 0.1 \\
Cyg X-1 distance & $d$ & kpc & 2.22 & 2.22\\
Axis position angle & $\psi$ & deg & 0 & 0\\
{\tt XILLVER} metal abundance relative to solar & $A_{\rm Fe}$ & none & 1 & 1\\
{\tt XILLVER} electron temperature & $T_{\rm e}$ & keV & 100 & 150\\
{\tt XILLVER} $e^{-}$-density in cm$^{-3}$ 
& $\log_{10}(n_{\rm e})$ & none & 17.5 & 17.7\\ 
Equivalent hydrogen column density & $N_{\rm H}$ & $10^{22}$ cm$^{-2}$ & 
0.2 & 4 \\ \hline
\end{tabular}
\label{t:kerrc}
\end{center}
\end{table}

\renewcommand{\thetable}{{\bf S4}}
\begin{table*}
\caption{\textbf{Optical polarization of Cyg X-1.} Normalized Stokes parameters $q$ and $u$ are presented for the observed polarization of the source ($q_{\rm obs}$, $u_{\rm obs}$), the IS polarization ($q_{\rm is}$, $u_{\rm is}$), and the intrinsic polarization obtained by subtracting the IS polarization from the observed values ($q_{\rm int}$, $u_{\rm int}$). The polarization degree (PD) and polarization angle (PA) of the intrinsic polarization are computed using formulae (\ref{eq:PDPA}).
Uncertainties are 1$\sigma$. }

\centering
\begin{footnotesize}
\begin{tabular}{p{2.5cm}|cc|cc|cc}
\hline
\hline
 Band & \multicolumn{2}{|c|}{$B$} & 
\multicolumn{2}{|c|}{$V$} & \multicolumn{2}{|c}{$R$} \\ \hline
   & $q$ (\%) & $u$ (\%) &  $q$ (\%) & $u$ (\%) &  $q$ (\%) & $u$ (\%) \\
\hline
\multicolumn{7}{c}{Observed polarization of Cyg~X-1} \\ \hline
DIPol-2 & $\phantom{-}0.46 \pm 0.06 $ & $-4.89 \pm 0.04$ & $\phantom{-}0.39 \pm 0.04$ & $-4.57 \pm 0.04$ & $\phantom{-}0.26 \pm 0.03$ & $-4.44  \pm 0.03$\\
RoboPol & -- & --  & --  & --  & $\phantom{-}0.61 \pm 0.13$ & $-4.74  \pm 0.12$\\
\hline
\multicolumn{7}{c}{Interstellar polarization} \\ \hline
Ref 2/DIPol-2  & $-0.09 \pm 0.17 $ & $-4.31 \pm 0.17$ & $-0.12 \pm 0.14$ & $-3.91 \pm 0.14$ & $-0.19 \pm 0.15$ & $-3.82  \pm 0.15$\\
Ref 1+2/DIPol-2  & $-0.41 \pm 0.11 $ & $-4.39 \pm 0.11$ & $-0.33 \pm 0.10$ & $-3.92 \pm 0.10$ & $-0.67 \pm 0.07$ & $-4.05  \pm 0.07$\\ 
Ref 2/RoboPol  & -- & --  & --  & --  &  $\phantom{-}0.39 \pm 0.16$ & $-4.00  \pm 0.08$\\
\hline
\multicolumn{7}{c}{Intrinsic polarization of Cyg~X-1\phantom{-}} \\ \hline
Ref 2/DIPol-2  & $\phantom{-}0.55 \pm 0.17$ & $-0.58 \pm 0.17$  & $\phantom{-}0.51 \pm 0.14$ & $-0.66 \pm 0.14$ & $\phantom{-}0.45 \pm 0.15$ & $-0.62 \pm 0.15$ \\ 
Ref 1+2/DIPol-2  & $\phantom{-}0.87 \pm 0.11$ & $-0.50 \pm 0.11$ & $\phantom{-}0.72 \pm 0.10$ & $-0.65 \pm 0.10$ & $\phantom{-}0.93 \pm 0.07$ & $-0.39 \pm 0.07$\\
Ref 2/RoboPol & -- & --  & --  & --  & $\phantom{-}0.22 \pm 0.21$ & $-0.74 \pm 0.14$ \\ 
\hline 
\hline
\multicolumn{7}{c}{Intrinsic polarization of Cyg~X-1} \\ \hline 
& PD (\%) & PA (deg) & PD (\%) & PA (deg) &  PD (\%) & PA (deg) \\
\hline
Ref 2/DIPol-2 & $\phantom{-}0.79 \pm 0.17$ & $-23 \pm 6$ & $\phantom{-}0.83 \pm 0.14$ & $-26 \pm 5$ & $\phantom{-}0.77 \pm 0.15$ & $-27 \pm 6$ \\
Ref 1+2/DIPol-2 & $\phantom{-}1.00 \pm 0.11$ & $-15 \pm 3$ & $\phantom{-}0.97 \pm 0.10$ & $-21 \pm 3$ & $\phantom{-}1.01 \pm 0.07$ & $-11 \pm 2$\\
Ref 2/RoboPol & -- & --  & --  & --  &  $\phantom{-}0.77 \pm 0.15$ & $-37 \pm 6$ \\
\hline 
\end{tabular}
\end{footnotesize}
\label{tab:dipol}
\end{table*}

\renewcommand{\thetable}{{\bf S5}}
\begin{table}[ht]
\caption{{\bf Best-fit parameters of the spectro-polarimetric model fitted to the data (Equation \ref{eqn:specpol})} Other \textsc{xillverCp} parameters were tied to the corresponding \textsc{relxillCp} parameters. The \textsc{relxillCp} reflection fraction has been multiplied by 15.043 to account for \textsc{nthComp} and \textsc{relxillCp} being normalized differently. 
The uncertainties are given on the 90\% confidence level. 
} 
\begin{center}
\begin{tabular}{p{2.5cm}p{4cm}p{5.3cm}p{2cm}}
\hline
\hline
Component & Parameter (unit) & Description & Value\\
\hline
\hline
\textsc{TBabs} & $N_{\rm H}$ ($10^{22}$\,cm$^{-2}$) & Hydrogen column density & $0.437^{+0.025}_{-0.10}$\\
\hline
\multirow{2}{2.5cm}{\textsc{diskbb}} & $kT_{\rm d}$ (keV) & Peak disk temperature & $0.319^{+0.018}_{-0.018}$ \\
                                   & \texttt{norm} ($10^3$) & Normalization & $3.79^{+0.90}_{-1.3}$ \\
\hline
\multirow{5}{2.5cm}{\textsc{nthComp}} & $\Gamma$ & Photon index & $1.62^{+0.0043}_{-0.0078}$\\
                                    & $kT_{\rm e}$ (keV) & Electron temperature & $94.2^{+2.4}_{-6.8}$ \\
                                    & \texttt{norm} & Normalization & $0.945^{+0.050}_{-0.044}$ \\
                                    & PD (\%) & polarization degree & $3.63^{+0.26}_{-0.26}$ \\
                                    & PA (deg) & polarization angle & $-20.5^{+2.1}_{-2.1}$ \\
\hline
\multirow{6}{2.5cm}{ \textsc{relxillCp} } & $r_{\rm in}$ ($r_g$) & Disk inner radius & $3.35^{+0.62}_{-0.41}$ \\
                                        & $i$ (deg) & Disk inclination angle & $37.8^{+1.2}_{-2.9}$ \\
                                        & $\log_{10}(\xi/[{\rm erg}~{\rm cm}~{\rm s}^{-1}])$ & Ionization parameter & $3.15^{+0.040}_{-0.031}$ \\
                                        & $kT_{\rm e}$ (keV) & Electron temperature & $140^{+32}_{-42}$ \\
                                        & $A_{\rm Fe}$ (solar) & Iron abundance & $3.70^{+0.50}_{-0.21}$ \\
                                        & $f$ (\%) & Reflection fraction & $20.17^{+1.6}_{-2.8}$ \\
\hline
\multirow{2}{2.5cm}{\textsc{xillverCp}} & $\log_{10}(\xi/[{\rm erg}~{\rm cm}~{\rm s}^{-1}])$ & Ionization parameter & $2.25^{+0.099}_{-0.19}$ \\
                                   & \texttt{norm} ($10^{-3}$) & Normalization & $3.46^{+0.14}_{-0.72}$ \\
\hline
\multirow{2}{2.5cm}{\textsc{mbpo} \\ \footnotesize{{NuSTAR} FPMA}} & $\Delta\Gamma_1$ ($10^{-2}$) & Power-law index & $-6.22^{+0.60}_{-0.67}$ \\
                                                & $N$ & Normalization & $1.13^{+0.0081}_{-0.0038}$ \\
\hline
\multirow{2}{2.5cm}{\textsc{mbpo} \\ \footnotesize{{NuSTAR} FPMB}} & $\Delta\Gamma_1$ ($10^{-2}$)  & Power-law index & $-7.11^{+0.56}_{-0.61}$ \\
                                                & $N$ & Normalization & $1.17^{+0.0051}_{-0.0034}$ \\
\hline
\multirow{2}{2.5cm}{\textsc{mbpo} \\ \footnotesize{{INTEGRAL}}} & $\Delta\Gamma_1$ ($10^{-2}$)  & Power-law index & $-13.8^{+2.2}_{-1.7}$ \\
                                                & $N$ & Normalization & $1.44^{+0.018}_{-0.075}$ \\
\hline
\multirow{4}{2.5cm}{\textsc{mbpo} \\ \footnotesize{{IXPE} DU1}} & $\Delta\Gamma_1$ ($10^{-2}$)  & Low energy power-law index & $1.34^{+1.8}_{-1.8}$ \\
                                                & $\Delta\Gamma_2$ ($10^{-2}$)  & High energy power-law index & $-22.2^{+1.3}_{-1.3}$ \\
                                                & $E_{\rm br}$ (keV) & Break energy & $3.28^{+0.13}_{-0.10}$ \\
                                                & $N$ & Normalization & $1.51^{+0.0071}_{-0.0084}$ \\
\hline
\multirow{4}{2.5cm}{\textsc{mbpo} \\ \footnotesize{{IXPE} DU2}} & $\Delta\Gamma_1$ ($10^{-2}$)  & Low energy power-law index & $-5.61^{+1.7}_{-1.4}$ \\
                                                & $\Delta\Gamma_2$ ($10^{-2}$)  & High energy power-law index & $-27.9^{+1.6}_{-2.2}$ \\
                                                & $E_{\rm br}$ (keV) & Break energy & $3.54^{+0.19}_{-0.15}$ \\
                                                & $N$ & Normalization & $1.45^{+0.0011}_{-0.013}$ \\
\hline
\multirow{4}{2.5cm}{\textsc{mbpo} \\ \footnotesize{{IXPE} DU3}} & $\Delta\Gamma_1$ ($10^{-2}$)  & Low energy power-law index & $-8.82^{+1.6}_{-1.7}$ \\
                                                & $\Delta\Gamma_2$ ($10^{-2}$)  & High energy power-law index & $-29.5^{+3.1}_{-3.3}$ \\
                                                & $E_{\rm br}$ (keV) & Break energy & $3.33^{+0.18}_{-0.15}$ \\
                                                & $N$ & Normalization & $1.44^{+0.012}_{-0.014}$ \\
\hline
\end{tabular}
\label{t:Spectrafitparams}
\end{center}
\end{table}
\end{document}